\documentclass[fleqn,usenatbib,useAMS]{mnras}

\pdfoutput=1 
\usepackage{amssymb}
\usepackage{amstext}
\usepackage{amsmath}
\usepackage{epsf}
\usepackage{graphicx}
\usepackage{graphics}
\usepackage{epsfig}
\usepackage{wrapfig}
\usepackage{listings}
\usepackage{float}
\usepackage{esint}
\usepackage{natbib}
\usepackage{ifthen}
\usepackage{hyperref}
\usepackage{verbatim}

\usepackage{xcolor}
\usepackage[normalem]{ulem} 
\newcommand\hl{\bgroup\markoverwith
  {\textcolor{yellow}{\rule[-.5ex]{2pt}{2.5ex}}}\ULon}

\newcommand\rhl{\bgroup\markoverwith
  {\textcolor{orange}{\rule[-.5ex]{2pt}{2.5ex}}}\ULon}
  
 \newcommand\ghl{\bgroup\markoverwith
  {\textcolor{green}{\rule[-.5ex]{2pt}{2.5ex}}}\ULon} 
\begin{document}
\label{firstpage}
\pagerange{\pageref{firstpage}--\pageref{lastpage}}

\title[Topology and geometry of the dark matter web]{Topology and geometry of the dark matter web: a multistream view}

\author[Ramachandra \& Shandarin]
	{Nesar S. Ramachandra, \thanks{E-mail: nesar@ku.edu} 
	Sergei F. Shandarin, \\
	Department of Physics and Astronomy, University of Kansas, Lawrence, KS 66045}

\maketitle
\begin{abstract}

Topological connections in the single-streaming voids and multistreaming filaments and walls reveal a cosmic web structure different from traditional mass density fields. A single void structure not only percolates the multistream field in all the directions, but also occupies over 99 per cent of all the single-streaming regions. Sub-grid analyses on scales smaller than simulation resolution reveal tiny pockets of voids that are isolated by membranes of the structure. For the multistreaming excursion sets, the percolating structure is significantly  thinner than the filaments in over-density excursion approach.  

Hessian eigenvalues of the multistream field are used as local geometrical indicators of dark matter structures. Single-streaming regions have most of the zero eigenvalues. Parameter-free conditions on the eigenvalues in the multistream region may be used to delineate primitive geometries with concavities corresponding to filaments, walls and haloes.

\end{abstract}

\begin{keywords}
methods: numerical -- cosmology: theory -- dark matter -- large-scale structure of Universe 
\end{keywords}

\begingroup
\let\clearpage\relax
\endgroup
\newpage

\section{Introduction} 
\label{sec:intro}

Large scale structures with  highly anisotropic shapes were first theoretically predicted by Zeldovich approximation (hereafter ZA) \citep{Zeldovich1970}. The model based on ZA suggested that the eigenvalues of the deformation tensor dictate the shapes of the {\it collapsed} structures at the beginning non-linear stage of gravitational instability (\citealt{Arnold1982}, see also \citealt{Shandarin1989} and \citealt{Hidding2014}). These structures were found to be crudely  characterised as two-, one- and zero- dimensional  which actually meant that three characteristic scales of each structure ($L_1\ge L_2\ge L_3$) are approximately related as  $L_1^{(p)} \approx L_2^{(p)} \gg L_3^{(p)}$ or $L_1^{(f)} \gg L_2^{(f)} \approx L_3^{(f)}$ or $L_1^{(h)} \approx L_2^{(h)}  \approx L_3^{(h)} $ respectively.  In addition it implied that $L_1^{(p)} \approx L_1^{(f)}$ and $L_3^{(p)} \approx L_2^{(f)} \approx L_1^{(h)}$.\footnote{The multi-scale character of the cosmic web was not discussed until 1990s.} At present these generic types of structures are referred to as  walls/pancakes/sheets/membranes, filaments and haloes. Although the accuracy of the Zeldovich approximation deteriorates from pancakes to  filaments and especially to halos on qualitative level  there are no more types of  structures. Altogether these structures contain the most of mass in the universe nevertheless they occupy  very little space. The most of space is almost empty  and is referred to as voids.

\cite{Klypin1983a} (firstly reported  in \citealt{Shandarin1983}) were the first to identify a `three dimensional web structure' in the N-body simulation of the hot dark matter scenario. The simulation with $32^3$ particles used Cloud-in-Cell (CIC) technique on equal mesh revealed  that the gravitationally bound clumps of mass -- haloes in the present-day terminology --  were linked by the web of filamentary enhancements of density which spanned throughout the entire simulation box with the side of about 150$h^{-1}$Mpc in co-moving space. In addition \cite{Klypin1983a}  suggested that pancakes must be considerably less dense than the filaments since they were not detected in the simulation. These  results were quickly confirmed by \cite{Centrella1983} and \cite{Frenk1983}. In addition \cite{Centrella1983} who ran the simulation on similar mesh but with 27 times more particles also detected pancakes at $\rho/\bar{\rho} = 2$ level. At present this picture is widely accepted, and is referred to as the `cosmic web' (\citealt{Bond1996} and \citealt{Weygaert2008c}). 

Galactic distributions in redshift surveys have also revealed distinct geometries and topologies of the cosmic web. One of the first indications of the connection of the clusters of galaxies by filaments was demonstrated by \cite{Gregory1978} who discovered a conspicuous chain of galaxies between Coma and A1367 clusters using a sample of 238 galaxies. Later  this result was confirmed by \cite{DeLapparent1986} who used a significantly greater redshift catalogue of 1100 galaxies of the same region. \cite{Zeldovich1982} compared the percolation properties of the redshift catalogue of 866 local galaxies provided by J. Huchra with three theoretical distribution of particle in space: a Poisson distribution, the  hierarchical model by \cite{Soneira1978} and the particle distribution obtained from N-body simulation by \cite{Klypin1983a}.  They found that the both the galaxy sample and the density field obtained in N-body simulation percolated at considerably smaller filling factors  than  the Poisson distribution. On the other hand the  hierarchical model percolated at higher filling factors  than  the Poisson distribution. Further studies confirmed that the galaxies and the particles in the hot dark matter  model are arranged in the web-like structures \cite{Zeldovich1982}, \cite{Shandarin1983}, \cite{Shandarin1983b}, \cite{Shandarin1984}. This result was confirmed in more detailed analysis by \cite{Einasto1984}. \cite{Melott1983b} also found similar percolation properties in the mass distribution in the N-body simulation of a CDM model.
 
Thus by the  early 1990s it was clearly demonstrated that the web like structure is a generic type for a wide range of initial conditions in both two-(\citealt{Melott1990}, \citealt{Beacom1991}) and three- dimensional \citep{Melott1993} cosmological N-body simulations. However it also was demonstrated that the quantitative parameters of the web structures depend on the initial  power spectrum. Remarkably the simulations also showed that  adding small scale perturbations does not ruin the large scale structures if the slope of the power spectrum is negative in both two- and three- dimensional simulations.

All aspects of these studies have been experiencing great advancements in  three decades passed since the discovery and first studies of the geometry and topology  of the large-scale structures. The galaxy redshift catalogues have grown by thousands of times (by surveys such as Sloan Digital Sky Survey (SDSS) \citealt{Tegmark2003} and \citealt{Albareti2016} and the 2MASS Redshift Survey \citealt{Huchra2012}), the sizes of cosmological N-body simulations (modern large scale simulations like Millennium \citealt{Springel2005b} and Q-Continuum \citealt{Heitmann2015}) by more than a million times. The number of various methods for identifying  structures has also grown practically from  one method\footnote{FOF was used for the topological studies via percolation technique and identifying super clusters of galaxies (\citealt{Zeldovich1982}, \citealt{Shandarin1983}, \citealt{Shandarin1983b} on the one hand and for identifying halos \citealt{Davis1985} on the other.}  to several dozens (\citealt{Colberg2008}, \citealt{Knebe2011a}, \citealt{Onions2012}, \citealt{Knebe2013} and references therein). Measuring or quantifying  the structures always has  been a difficult problem and many sophisticated  techniques both mathematically and computationally have been proposed and investigated (see reviews by \citealt{Weygaert2008c}, \citealt{Weygaert2008}).

Cosmic web structures have been characterized using several geometrical and topological indicators such as genus curves (\cite{Gott1986}). In an attempt to characterize the shapes of  individual regions in the excursion sets of the density field, \cite{Sahni1998} suggested to use partial Minkowski functionals. They developed the method labelled SURFGEN and applied it to CIC density field obtained in N-body simulations 
(\citealt{Sathyaprakash1998}, \citealt{Sheth2003}, \citealt{Shandarin2004}). \cite{Aragon-Calvo2007} have developed the multi-scale MMF (Multi-scale Morphology Filter) detection technique based on the signs of three eigenvalues of the Hessian computed for  a set of replicas of the density field filtered on different scales. Similar multi-scale approaches to identifying structures is adopted in NEXUS and its extensions to velocity shear, divergence, and tidal fields \cite{Cautun2013}. More recently, persistence and Morse-Smale complexes in the density fields are analysed by \cite{Sousbie2011a}, \cite{Sousbie2011b} and \cite{Shivshankar2015a} to detect multi-scale morphology of the cosmic web.


There is also an increasing interest in the measures for detecting filaments in large astronomical surveys. Topology in the large scale structure was analysed by Betti Numbers for Gaussian fields \citep{Park2013} and SDSS-III Baryon Oscillation Spectroscopic Survey \citep{Parihar2014}. \cite{Sousbie2008a} detected skeleton of filaments of the SDSS and compared to the corresponding galaxy distribution. In smoothed density of mock galaxy distribution, \cite{Bond2010a} studied the projection of eigenvalues. The Hessian eigenvector corresponding to the largest eigenvalue is used by \cite{Bond2010b} to trace individual filaments in N-body simulations and the SDSS redshift survey data. Majority of the above analyses, however, ignore the dynamical information from the velocity field.



On the other hand, detection of voids and study of their morphological properties are done via numerous methods too. Traditional detection of void regions using just the particle coordinates differ based on the various methods used to identify them (see comparison of void finders in \citealt{Colberg2008} and references therein). Some methods involve using under-density thresholds. \cite{Blumenthal1992} proposed that the mean density in voids is $\delta = -0.8$ by applying linear theory argument. Similar threshold was used by \cite{Colberg2005} to identify voids. Under-dense excursion set approach was used by \cite{Shandarin2006} to identify percolating voids. \cite{Sheth2004a} used the excursion set formalism to develop an analytical model for the distribution voids in hierarchical structure formation (also see the excursion set approaches applied to voids by \citealt{Paranjape2012}, \citealt{Jennings2013} and \citealt{Achitouv2015}). Voids are also detected by isolating regions around local minima of density fields. For instance, the watershed transform is used by WVF-\cite{Platen2007},  ZOBOV-\cite{Neyrinck2008} and VIDE-\cite{Sutter2015} for segmentation of under-dense regions.

The unfiltered density field was generated using DTFE-Delaunay Tessellation Field Estimator (\citealt{Schaap2000}, \citealt{Weygaert2009a} and \citealt{Cautun2011}) by applying it to the particle coordinates. Earlier it was shown that DTFE is superior to CIC techniques (\citealt{Schaap2007} and \citealt{Weygaert2009a}) in generation of the density field with high spatial resolution. In a new approach to the analysis of the shapes of the large-scale structures, \cite{Sousbie2011a} introduced DIScrete Persistent Structure Extractor ({DisPerSE}) based on Morse-smale complex. By implementing it on realistic cosmological simulations and observed redshift catalogues \cite{Sousbie2011e} found that DisPerSE traces very well the observed filaments, walls and voids.

An additional dimension to the scope of the structure shapes is related to the question whether the density distribution (regardless of it form: continuous or discrete) is the only physical diagnostic of the cosmic web shapes or not. If not, then whether it is the best of all or not. And even if it is the best, then whether the other fields or distributions can provide a valuable contribution to understanding the shapes of the cosmic web or not. The answer to the latter question seems to be positive. In fact there are examples of attempts to bring new players into the field. For instance \cite{Hahn2007} and \cite{Forero-Romero2009a} studied the relation between the geometry of structures and the Hessian of the gravitational potential. \cite{Shandarin2011} demonstrated that the study of the multistream field reveals some features of the structures that cannot be easily seen in the density field. This has become even more evident when \cite{Shandarin2012} and \cite{Abel2012b} showed that the full dynamical information in the form of three-dimensional  sub-manifold in six-dimensional phase space can be easily obtained from the  initial and final coordinates of the particles in DM simulations. \cite{Hahn2015a} showed that this method provides extremely accurate estimates of the cosmic velocity fields and its derivatives. It has been shown that the multistream field provides a physical definition of voids in N-body DM simulations by the local condition $n_{str} = 1$ (\citealt{Shandarin2012} and \citealt{Ramachandra2015}). \cite{Falck2012} proposed the {ORIGAMI} method of assigning particles to  structures based on the number of axes along which particle crossing has occurred. Void, wall, filament, and halo particles are particles that have been crossed along 0, 1, 2, and 3 orthogonal axes, respectively. \cite{Shandarin2016} identify the void particles as the ones that do not undergo any {\it flip-flop} through the evolution. Each of above definitions completely independent of any free parameters, with small differences in the physical implication.

Tracing the Lagrangian sub-manifold also provides rich insights into caustics (\cite{Arnold1982} and \citealt{Hidding2014}) and halo collapse \cite{Neyrinck2015a}. Recently, there are attempts to improve N-body simulations (see \cite{Hahn2013}, \cite{Angulo2013a}, \cite{Angulo2013b}, \cite{Sousbie2015} and \cite{Hahn2016a}) by solving the Vlasav-Poisson equation using tessellations in the Lagrangian sub-manifold. Galaxy evolution and star formation in the context of multi streaming phenomenon are studied by \cite{Aragon-Calvo2016b}. 


Despite the considerable improvements  in simulating, identifying and measuring the cosmic web 
-- briefly discussed above -- many aspects remain unsettled and are vigorously debated. 
The intention of this work is to further investigate the strengths and weaknesses of the multistream field as a complimentary diagnostic of the shapes in the DM web. 
Multi-stream filed is simply the number of DM streams at every point of Eulerian space.  Thus it is an odd positive integer at a given point (\citealt{Arnold1982}, see also \citealt{Shandarin1989} and \citealt{Hidding2014}). We estimated it  on a regular mesh of a chosen resolution from the tessellation of of the simulation particles in Lagrangian space and the particle coordinates at a chosen time \cite{Shandarin2012}.
The external boundaries of the cold DM web are the caustics in the density field which are clearly seen in
the simulations with adequate resolution of the density field (see e.g. Fig 7 in \citet{Hahn2015a}). However the exactly same boundaries of the DM web can be identified as the boundaries of a single-stream flow which is a local parameter. The multistream field even a better indicator of the boundaries of the DM web than caustics because caustics are present everywhere  the number of streams varies (from 1 to 3, from 3 to 5, etc) but the boundary of the web are only the one where the number of stream changes from 1 to 3.

In particular we would like to discuss the differences in defining voids in density and multistream fields. It is closely related to the definition and distinguishing of linear and non-linear structures or regimes. One simple statistical definition that often used  is as follows:  after defining the std of the density contrast $\sigma_{\delta} \equiv <(\rho(x)/ \bar{\rho} - 1)^2>^{1/2}$ one can roughly separate the linear and non-linear regimes by the boundary $\sigma_{\delta}=1$. This is obviously very crude characteristic which does not say much about the geometry and topology of the non-linear structures. The parameter $\sigma_{\delta}$ is frequently is evaluated for filtered fields $\sigma_{\delta} = \sigma_{\delta}(R_{\rm f})$. Unfortunately the transition from `non-linear'  field at small $R_{\rm f}$ to `linear' field at large  $R_{\rm f}$ is smooth and thus choosing a particular value of  $R_{\rm f}$ is remarkably subjective. 

A related but different question is how to select individual non-linear structure, like halos, filaments and walls by using a local parameter. In particular the density threshold has been  used on numerous occasions especially for identifying halos and voids. As a rule the choices of particular values have not been justified by solid physical evidences. The virial mass and virial radius of a halo are often used as direct indicators of gravitationally bound objects but they are  determined  by a nonlocal quantity -- the mean overdensity of the halo. 
An interesting compaison of several kinds of boundaries of halos was provided by \cite{More2015}. In particular they considered  the virial radius $R_{\rm vir}$,  $R_{\rm  200m}$,  the splashback radius $R_{\rm sp}$, and $R_{\rm infall}$. The splashback
radius is defined as an average distance from the center of the halo to the most external caustic if it was resolved. The authors argue
that it is ``a more physical halo boundary choice" than ``commonly defined to enclose a density contrast $\Delta_{\rm m,c}$ relative to 
a reference (mean or critical) density.
This is the boundary where the number of streams falls from three to one in the multistream field.

Gravitationally bound structures could be defined as linear in the sense  that $\delta({\bf x}) \ll 1$ for all points in the structure. A simple example is a progenitor of large halo at linear stage. However  one cannot accurately identify such an object at linear
stage using a local criterion like a density threshold. Even at the nonlinear stage of N-body simulation one cannot predict 
when a particular fluid element with a given value of $\delta$ in a void will be accreted to a wall or filament. Among other factors 
the size of the void and proximity to a wall would play significant roles.  In addition the walls accrete expanding fluid elements
as well thus the velocity divergence on the fluid element would not help.


The rest of the paper is organised as follows: we describe the cosmological simulations in Section \ref{sec:simulation}. Some of the important features of the multistream field are described in Section \ref{sec:multiCalc}. Topology of the single-streaming voids is discussed in \ref{sec:voids} and that of the multistream structure is investigated using percolation theory in Section \ref{sec:percolation}. . Discussion of the local geometry of multistream field using Hessian matrices is done in Section \ref{sec:hessian}. 

\section{The simulation}
\label{sec:simulation}

In this analysis, we use cosmological N-body simulations generated by the tree-PM code {GADGET-2}    (\citealt{Springel2005} and \citealt{Springel2001}). 
The periodic side lengths $L$, number of particles $N_p$, masses of each particle $m_p$ and the gravitational softening length $\epsilon$ for the two simulations are tabulated in \autoref{tab:Simulations}. Initial conditions at redshift of $z_{ini}= 80$ are generated by {MUSIC} \citep{Hahn2011a} with the transfer function from \cite{Eisenstein1998a}. We adopt the $\Lambda$CDM cosmological model with cosmological parameters $\Omega_{m}= 0.276$, $\Omega_{\Lambda}= 0.724$, the Hubble parameter, $h = 0.703$, the power spectrum normalization, $\sigma_8 = 0.811$ and the spectral index $n_s= 0.961$. 

\begin{table}
  \caption{Parameters for the simulation boxes: Side length $L$, number of particles $N_p$, mass of each particle $m_p$, and the gravitational softening length $\epsilon$ for the GADGET simulations are shown.}
\begin{tabular}{l|c|c|c| }
\hline
$L$ & $N_p$ & $m_p$ & $\epsilon$  \\  \hline
$100 h^{-1} Mpc$ & $128^3$      & $ 3.65 \times 10^{10} h^{-1} M_{\sun}$  & $20 h^{-1} kpc$\\ \hline
$100 h^{-1} Mpc$ & $256^3$      & $ 4.57 \times 10^{9} h^{-1} M_{\sun}$  & $10 h^{-1} kpc$ \\ \hline

\end{tabular}
\label{tab:Simulations}
\end{table}

\subsection{Multi-stream field at $z=0$}
\label{sec:multiCalc}

\begin{figure} 
\centering\includegraphics[width=8cm]{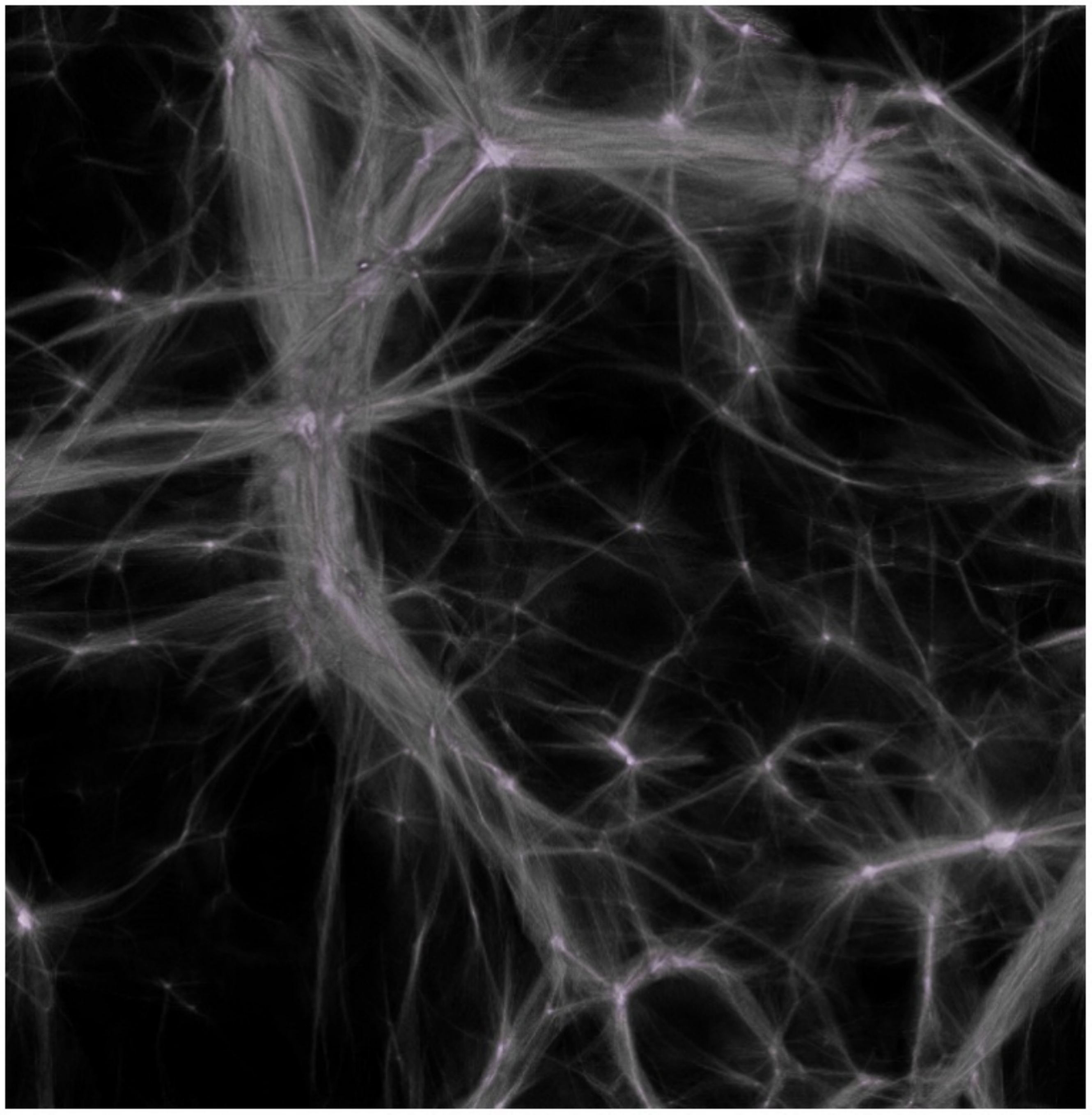} 
\caption{3D rendering of the multistream field: the cosmic web structure of a $ 50 h^{-1} \text{Mpc} \times 50 h^{-1} \text{Mpc} \times 50 h^{-1} \text{Mpc}$ slice in a simulation box of side length $100 h^{-1}$ Mpc and $128^3$ particles. The multistream field is calculated at 8 times the native resolution. void(black) is a percolating structure with $n_{str} = 1$. Regions $n_{str} \geq 17$ show a filamentary structure (gray) and the bright spots at the intersections of the filaments are regions with $n_{str} \geq 100$. }
\label{fig:full}
\end{figure}

The multistream field objectively characterizes the level of non-linearity in the cosmic web. The `number-of-streams' field or $n_{str}(\bmath{x})$ is computed from the Lagrangian sub-manifold $\bmath{x}(\bmath{q})$, which is a continuous three-dimensional sheet in a six-dimensional  $(\bmath{q}, \bmath{x})$ space. In this paper, we utilize the tessellation implementation by \cite{Shandarin2012} to calculate the multistream flow field on the GADGET-2 snapshot at $z=0$. This implementation only requires initial and final coordinates of the dark matter particles. 

The $n_{str}(\bmath{x})$ values are mostly odd-numbered since each folding in the Lagrangian sub-manifold results in an increase of $n_{str}$ by 2. Exception to this are only at caustics - which have volume measure zero, then the $n_{str}$ is even-valued number. The particles in $n_{str} = 1$ have not experienced orbit crossings and thus these regions are unambiguously identified as void \citep{Shandarin2012}. Foldings in the Lagrangian sub-manifold generally occur one-by-one. For example, a contour of $n_{str} = 7$ will be within a region of $n_{str} \leq 5$. Hence the multistream field commonly has nesting shells, i.e., $ 3 \supseteq  5 \supseteq  7 \supseteq  9 \supseteq  11 \ldots$. Some of the important features of the multistream field are discussed in Appendix \ref{appendix:nstream}.  

The first non-linear DM structures that reach non-perturbative stage of gravitational evolution have $n_{str} = 3$. By visual inspection, these regions generally form a fabric-like open structures that resemble walls. N-body simulations suggest that a DM fluid element after the first 
crossing of a caustic never returns in  a single-streaming state. Therefore the {\it local} condition $n_{str}({\bf r}_{\rm f.e.}) \geq 3$
(where  ${\bf r}_{\rm f.e.}$ is the position of the fluid element) is  sufficient for the fluid element to be bound to the DM web.

All particles that have fallen into a wall will never return to any single-streaming regions, therefore they can be labeled as gravitationally bound to pancakes/walls. The surface contours of higher $n_{str}$ are embedded within the walls. \autoref{fig:full} shows a filamentary structure of the multistream web at $n_{str} \geq 17$. The figure also shows regions around local maxima of the multistream field, which are generally located at the intersections of filaments.

The multistream field can be computed at arbitrary resolutions of diagnostic grids. The parameter `refinement factor' denotes the ratio of separation of the particles in Lagrangian grid, $l_l$, to side length of diagnostic grid $l_d$. In a simulation of $128^3$ particles, for instance, multistream field computed on a diagnostic grid of size $256^3$ would have a refinement factor of $l_l/l_d = 2$. 

\section{Voids in the multistream field}
\label{sec:voids}

\begin{figure*}
\begin{minipage}[t]{0.99\linewidth}
 \centering\includegraphics[height=11cm]{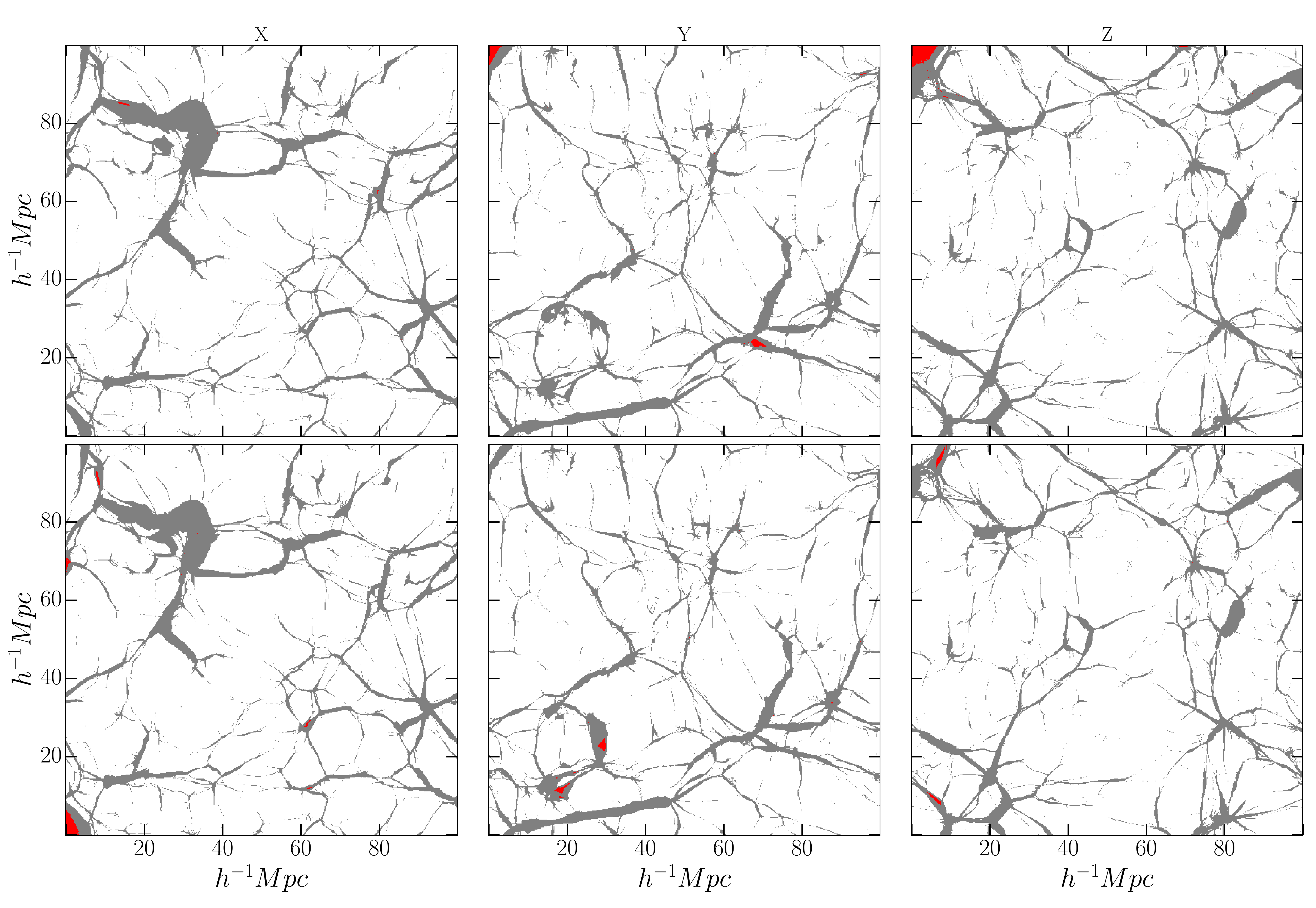} 
\end{minipage}\hfill
\caption{Opposite faces of the multistream field for the simulation box with $N_p = 128^3$. Non-void regions (gray) have $n_{str} > 1$. The largest void (white) in the entire field spans over the entire box. Rest of the smaller isolated voids (red) occupy very small volume fraction. }
\label{fig:voidFace}
\end{figure*}

Gravitational instability results in movement of the collision-less fluid particles in the Universe from voids to walls, walls to filaments, and filaments to haloes. As we mentioned above in the multistream portrait, the entry of mass particles from single-streaming regions into $n_{str} > 1$ region is irreversible. The converse is obviously not true, that is, the particles in $n_{str} = 1$ regions may move to multistreaming region at a later time in the evolution. At a given cosmic time, sufficient condition for dark matter particles to be bound to non-perturbative and non-linear structures like walls/filaments/haloes is being in multistream regions. Therefore, a single-stream flow implies that gravitationally bound structures haven't yet formed, and thus defined as a void region. This definition of void is unambiguous and physically motivated, as demonstrated by \cite{Shandarin2012}. It is worth stressing that while the density in voids varies, the number-of-streams is uniformly equal to unity.

For simulation box with $128^3$ particles, $n_{str} = 1$ regions have a large volume fraction of $VF_V \approx 93$ per cent regardless of the value of refinement factor (shown in \autoref{tab:VoidStat}). Multi-stream web structure in the simulation with higher mass resolution ($N_p = 256^3$) is better enhanced, and the single streaming void occupies around $90$ per cent of the volume. \autoref{fig:voidFace} shows the single streaming voids occupying large volume of the simulation with $128^3$ particles at refinement factor of 4. 

\begin{table}
  \caption{Volume fraction $VF_V$ of the voids, total number of isolated voids $N_V$ and the filling fraction of the largest void  $FF_1/VF_V$ at different refinement factors $l_l/l_d$. The filling fractions of the largest void at each refinement factor show that most of the $n_{str} = 1 $ region is almost entirely a single percolating structure.}
\begin{tabular}{|r|r|r|r|r| }
\hline
$N_p$ & $l_l/l_d$ & $VF_V$ & $N_V$ & $FF_1/VF_V$  \\  \hline
$ 128^3$      & $ 1$      & 93.46\%   & 1    & 100\% \\ \hline
$ 128^3$      & $ 2$      & 93.44\%   & 11   & 99.999\% \\ \hline
$ 128^3$      & $ 4$      & 93.44\%   & 113  & 99.999\% \\ \hline
$ 128^3$      & $ 8$      & 93.44\%   & 914  & 99.997\% \\ \hline
$ 256^3$      & $ 1$      & 90.80\%   & 11   & 99.999\% \\ \hline
$ 256^3$      & $ 2$      & 90.80\%   & 97   & 99.999\% \\ \hline
$ 256^3$      & $ 4$      & 90.80\%   & 1029 & 99.997\% \\ \hline
$ 256^3$      & $ 8$      & 90.80\%   & 7259 & 99.964\% \\ \hline

\end{tabular}
\label{tab:VoidStat}
\end{table}

\subsection{Connectivity of the voids}
\label{sec:voidPerc}

In order to find whether the void regions of the multistream field are connected or not, we isolate three-dimensional segments with $n_{str} = 1$ and separately label them. The number of disconnected voids in the simulation with $N_p = 128^3$ range from 1 (for refinement factor, $l_l/l_d =  1$) to about $900$ (for $l_l/l_d= 8$) as shown in \autoref{tab:VoidStat}. Number of isolated voids increases similarly in the simulation with $N_p = 256^3$ particles as well. 

Smoothing of the structure at lower resolution of the multistream field results in increased connectivity of single-streaming regions. In \autoref{fig:voidFace}, opposite faces on each axes of the multi-field, show a large connected void (white). This means that the largest void percolated throughout the multistream field in all directions. This result is in agreement with \cite{Falck2015}, who studied percolation of ORIGAMI-voids in simulations with side lengths of $100$ and $200  h^{-1} \text{Mpc}$. In addition to the percolating the field, the largest void also fills most of the void volume: the ratio of filling fraction of the largest void $FF_1$ to the volume fraction of $n_{str} = 1$ regions in the simulation is close to unity (see \autoref{tab:VoidStat}). This phenomenon is seen at each of the refinement factors in our analysis. Hence, over $99.9$ per cent of the single-streaming sites are connected throughout the simulation box, and they form a single empty region.

As previously mentioned, the multistream web structures of $n_{str} = 3$ form the first gravitationally collapsed structures. These tiny structures are better resolved in higher refinement factors, and they tend to enclose greater number of pockets of single-streaming voids inside them. The red regions in \autoref{fig:voidFace} some of the small voids on faces of the simulation box with $128^3$ particles. Despite increase in the number of small voids at each of the refinement factors, these void regions (i.e., the single streaming regions excluding the largest void) collectively occupy less than $0.1$ per cent of the total void volume in both the simulations. It is also likely that the small voids are simply due to numerical noise. However, the major conclusion regarding small voids remains the same up to refinement factor of 8. We do not pursue further investigation due to tiny effects.

\subsection{Halo boundaries within the void}
\label{sec:voidHaloes}

\begin{figure}
\begin{minipage}[t]{.99\linewidth}
\centering\includegraphics[width=8.cm]{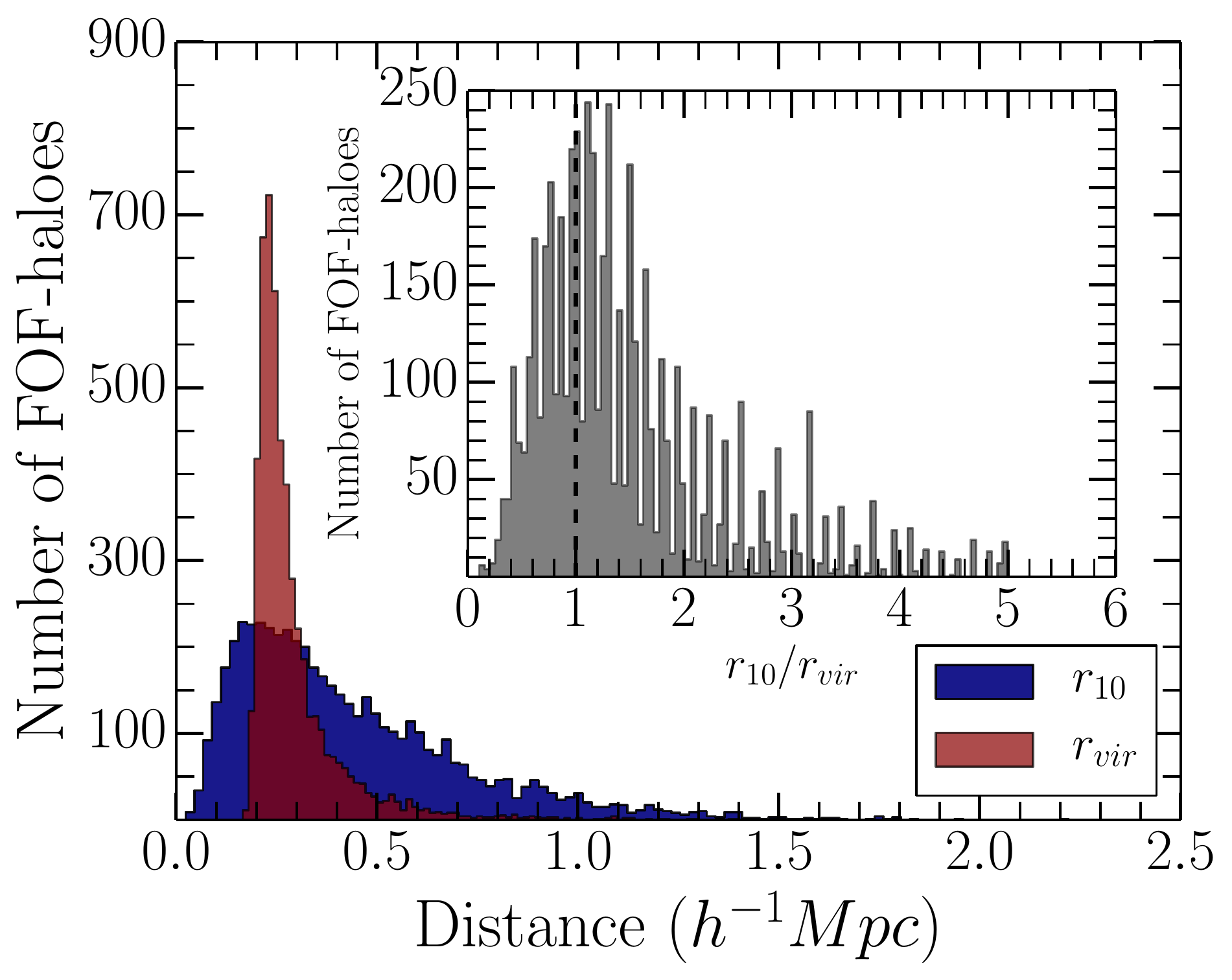} 
\end{minipage}\hfill
\caption{Single-streaming void distribution on diagnostic spheres around FOF-haloes are considered. At radius $r_{10}$, each diagnostic sphere has $n_{str} = 1$ on 10 per cent of its spherical surface. Distribution function of $r_{10}$ (blue) and FOF-radii $r_{vir}$ (red) are shown. Inner plot shows the distribution function of $r_{10}/r_{vir}$. The haloes within the dashed line have at least 10 per cent of their virial-surfaces in contact with $n_{str} = 1$ regions.}
\label{fig:Vfr_all}
\end{figure}

Dark matter haloes are the most non-linear objects in the cosmic web. With the exception of ORIGAMI \citep{Falck2012}, most of the halo finders do not consider multistreaming in the configuration space for finding haloes. Potential haloes found by several such halo finding methods, hence, may have boundaries that intersect with the single-streaming void, which is the least non-linear structure in the dark matter universe. \cite{Colberg2008} even mention existence of `void-haloes' in several halo finder algorithms.  

We studied the $n_{str}$ environment of the haloes detected using the Friends-of-Friends method (FOF-\citealt{Davis1985}) as illustrated in \autoref{fig:Vfr_all}. FOF-haloes with more than 20 particles are detected using linking-length of $b=0.2$ in the simulation with $128^3$ particles. We implement the diagnosis method prescribed in \cite{Ramachandra2015}: a large number of points are randomly selected on diagnostic spherical surfaces centred at the FOF-centre of the halo. Multi-stream values are iteratively calculated at these spherical surfaces of various radii. We define the distance from centre of a halo, $r_{10}$, where $n_{str} = 1$ at 10 per cent of the surface of the diagnostic sphere. Distribution of this void-distance parameter is compared to the virial radii $r_{vir}$ of the FOF-haloes. Surprisingly, $r_{10}$ distribution peaks at slightly lower values than the $r_{vir}$ distribution. This implies a large number of FOF-haloes are in the vicinity of the void. 

For specific examples of some FOF-haloes, \cite{Ramachandra2015} showed that single-stream may appear within their virial radii too. The distribution of $r_{10}/r_{vir}$ in the inner plot of \autoref{fig:Vfr_all} shows the same phenomenon. The FOF-haloes within $r_{10}/r_{vir} < 1$ (represented by the vertical dashed line) have $n_{str} = 1$ on 10 per cent of their virial surfaces. The figure illustrates that a large number of FOF-haloes satisfy this condition, thus are in contact with the void surfaces. Hence not all the FOF particles have undergone a gravitational collapse during their evolution.

For methods such as FOF, there is no unambiguous linking-length criterion for voids. Similarly for the density fields, a range of under-densities are prescribed by various void finder methods (cf. \citealt{Colberg2008}). On the other hand, the multistream field unambiguously identifies all the regions without a single gravitational collapse as voids. Haloes detected on the multistream field may address the issue of haloes being in contact with voids. 

\section{Percolation in the multistream web}
\label{sec:percolation}

\begin{figure}
\begin{minipage}[t]{.99\linewidth}
  \centering\includegraphics[width=8.cm]{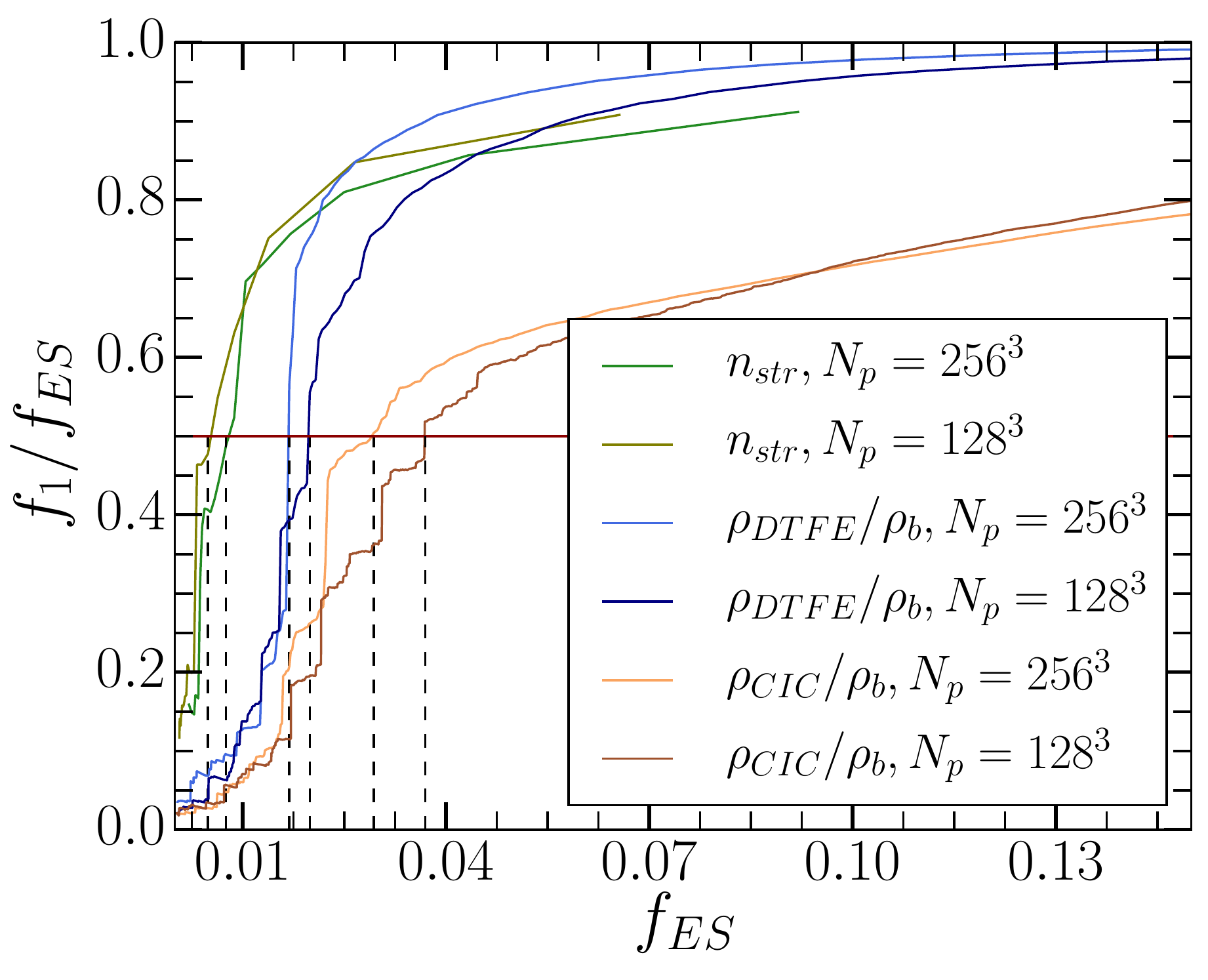} 
\end{minipage}\hfill
\caption{Percolation plot in the multistream field and mass density. Two density estimators - CIC and DTFE are shown. Percolation transition (at $f_1/f_{ES} = 0.5$ shown by the horizontal red line) occurs at smaller excursion set volumes for the multistream field, as seen by the dashed lines for both the curves. It is worth stressing that the percolation curves for $n_{str}$ field are bounded by conditions $f_{ES} < 0.1$. }
\label{fig:PercTh}
\end{figure}

\begin{figure*}
\begin{minipage}[t]{.99\linewidth}
  \centering\includegraphics[height=5.7cm]{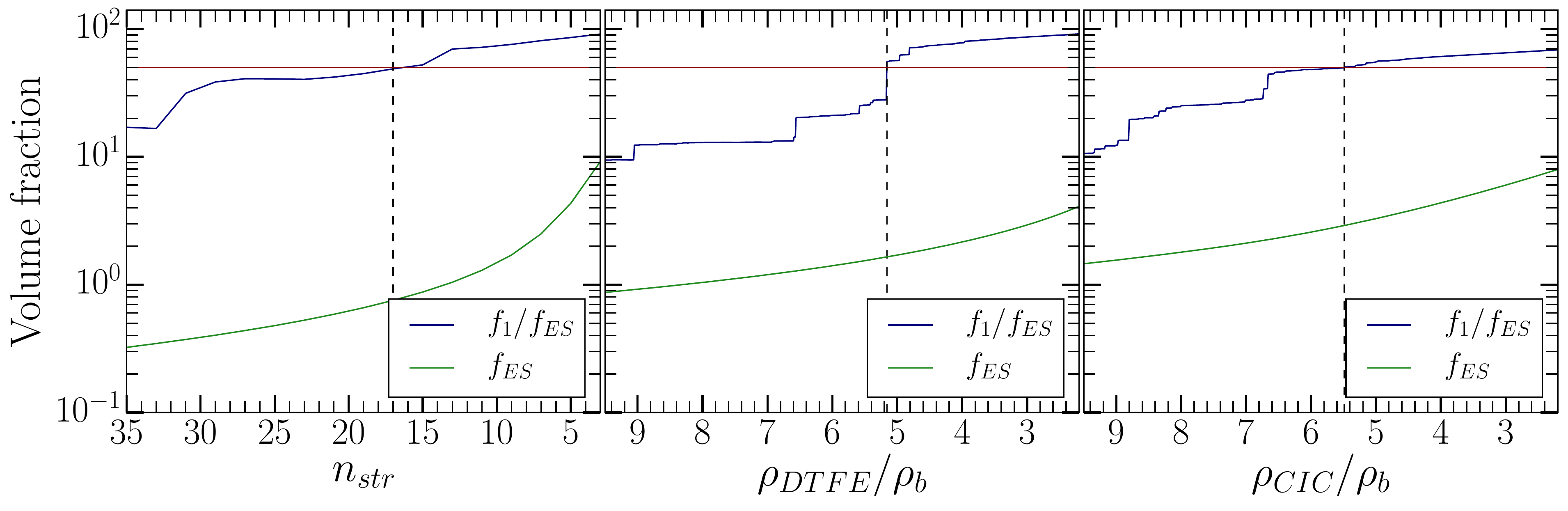} 
\end{minipage}\hfill
\caption{Percolation threshold in the multistream (left panel) and matter density fields. Matter density is calculated using DTFE (middle panel) and CIC (right panel) with a refinement factor of 2 in the simulation with $256^3$ particles. The volume fraction of excursion set and the filling fraction of the largest structure is shown. Percolation transition in multistream field at $n_{str} = 17$ is shows by the dashed vertical line. Percolation at $\rho_{DTFE}/ \rho_b = 5.16 $ and $\rho_{CIC}/ \rho_b = 5.49 $ are shown by the dashed vertical line. }
\label{fig:HaloFil}
\end{figure*}

A single percolating void fills the $n_{str} = 1$ regions almost entirely, as discussed in Section \ref{sec:voidPerc}. Disconnected pockets of void may exist, but they collectively occupy very small volume fraction (less than 0.1 per cent of the total volume as tabulated in \autoref{tab:VoidStat}). Whereas, the non-void structure in the multistream field has a different topological structure. The regions selected with a lower bound on $n_{str}$ could be isolated (generally for high $n_{str}$ thresholds) or connected in a percolating region (for low $n_{str}$ thresholds). We investigate the topological transitions in these excursion sets of multistream field.

The volume fraction as a function of number-of-streams decreases according to a power law in the $n_{str} > 1$ structure (\citealt{Shandarin2012} and \citealt{Ramachandra2015} report $\text{VF}(n_{str})$ decreasing as ${n_{str}}^{-2.8} $ and  ${n_{str}}^{-2.5}$ respectively for their simulations). The volume fraction of the excursion set $f_{ES}(n_i)$ is the ratio of volume of all the regions with a lower bound $n_i$ on the multistream field to the total volume $V_{tot}$ of the simulation box, i.e, $ \displaystyle f_{ES}(n_i) = \frac{V_{ES}}{V_{tot}} =  {\sum\limits_{n_{str} \geq n_i} \text{VF} ({n_{str}})}$. Since volume fraction of the each $n_{str}$ rapidly increases with an decrease in multistream value, so does the $f_{ES}$.

The excursion set may have number of isolated segments of different volumes. A measure of connectivity in the excursion set regions can be given by the filling fraction, $f_1/f_{ES}$, where $f_1$ is the volume fraction of the largest isolated region in the excursion set. $f_1$ can be computed numerically in the simulations. If the value of $f_1/f_{ES}$ is close to 0, then none of the isolated regions dominate the excursion set. This implies absence of percolation. If $f_1/f_{ES}$ is close to one, it implies a single connected structure dominates most of the excursion set.

The filling fraction $f_1/f_{ES}$ grows from 0 to 1 occurs rapidly $f_{ES}$ during percolation phase transition. A practical robust definition of the percolation transition is at $f_1/f_{ES} = 0.5$, i.e, when the  largest region occupies more than 50 per cent of the excursion set volume. The percolation plot in \autoref{fig:PercTh} reveals this phenomenon. Excursion volume fraction $f_{ES}$ at this transition, $f_{ES}^{(p)} = 0.48$ and $0.75$ per cent for the simulations with with $128^3$ and $256^3$ particles respectively (although the numbers were obtained in one simulation each. The difference may be well within the range of statistical errors for this size of simulation box). After the percolation transition, the filling fraction of the largest structure stabilizes towards unity.

The nature of the transition in mass density field is similar to that in multistream field. For the simulation simulation with $256^3$ particles, the density is calculated using CIC method at $256^3$ and $512^3$ grid points. In \autoref{fig:PercTh}, the percolation phenomenon in both mass density fields is shown along with that of multistream fields. The excursion set volume fraction at percolation transition, $f_{ES}^{(p)}$ is lower for multistream field, because the filaments in the multistream field are thinner than that of density picture. Volume fraction of the largest structure detected in the density field also tends to unity with decreasing $f_{ES}$, albeit less rapidly as that of the multistream field. This means that while the largest structure in a multistream web occupies most of the structure, the over-density excursion set is more fragmented.

The excursion volume fraction of the multistream web structure is limited to a small fraction of  less than 10 per cent since rest of the volume is void. The excursion set volume fraction increases with decreasing number-of-streams and reaches it's maximum at $n_{str} = 3$.  At this limit the filling fraction $f_1/f_{ES}$ is still less than unity, about 95 per cent. These two peculiar properties of the multistream field explain the shape of the percolation curves in \autoref{fig:PercTh}.
Since the multistream flow field is a discrete data field, the percolation transition is seen to occurs at a particular value of $n_{str}$ rather than a large range of values. For $n_{str} = 17$, the largest structure in the excursion set occupies more than half the volume of the entire excursion set. At this multistream threshold, the largest segment starts spanning large volume of the simulation box (as observed in the left panel of \autoref{fig:HaloFil}). The volume fraction of the excursion set at this percolation transition is $f_{ES}^{(p)} = 0.75$ per cent for simulation with $256^3$ particles. 

The percolation transition at $n_{str} = 17$ could be used as a criterion for detecting filaments in the cosmic web. Since the largest $n_{str} \geq 17$ region occupies more than 50 per cent of the excursion set, it is essentially the `backbone' of the cosmic web \citep{Shandarin2010b}. Heuristic analysis as discussed by \cite{Ramachandra2015} also arrived at the same threshold for identifying filaments. That analysis was based on a multistreams variation in halo environments, hence a local value. From our percolation analysis, we see that it is also justified globally.

In the simulation with $256^3$ particles, percolations in the density field occurs at $\rho_{DTFE}/ \rho_b = 5.16 $ and $\rho_{CIC}/ \rho_b = 5.49 $ for densities calculated with DTFE and CIC respectively. Here $\rho_b = 256^3 / 100^3 M_{\sun} h^{-3} \text{Mpc}^{-3}$, the background density. Notice that these values correspond to the density as calculated by the CIC and DTFE algorithms, and it might be different for other density finding methods. The volume fraction of the excursion set of over-densities at the percolation, $f_{ES}^{(p)} = 2.7$ per cent, is considerably higher than the corresponding $f_{ES}^{(p)}$ value in the multistream field. This implies that the percolation occurs at larger values of filling fraction in mass densities.


\section{Local geometry of the multistream field}
\label{sec:hessian}
\begin{figure}
\begin{minipage}[t]{.99\linewidth}
  \centering\includegraphics[width=8.cm]{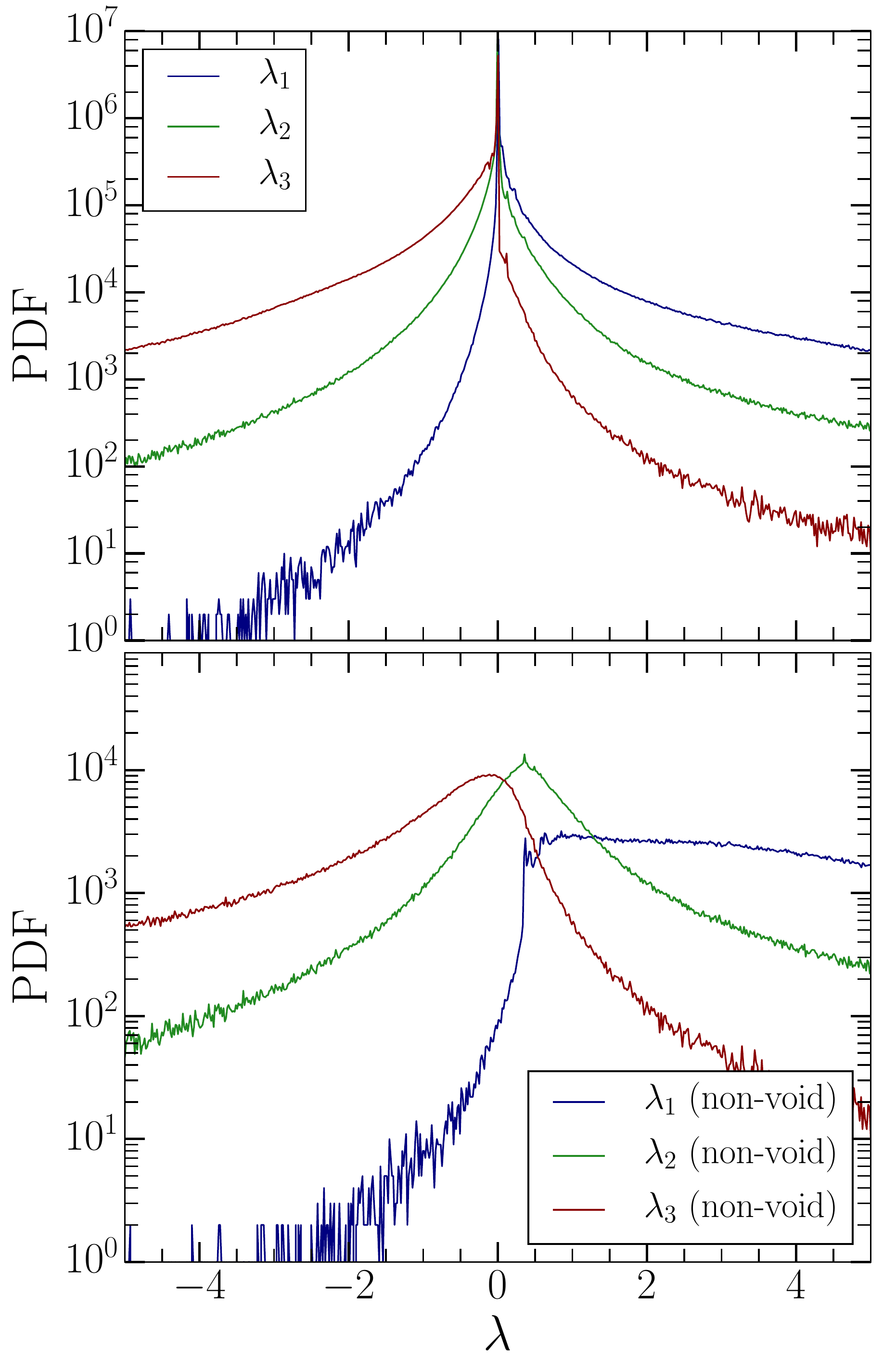} 

\end{minipage}\hfill
\caption{Probability distribution function of the sorted eigenvalues of the Hessian $\mathbfss{H}(-n_{str})$ in the simulation box with $N_p = 128^3$. Top panel: Distribution in the entire simulation box. The multistream field is calculated at refinement factor $l_l/l_d= 2$ and smoothing scale of equal to $l_d$. All the three eigenvalue data fields have a highest number of points where their value is 0. Bottom panel: Hessian eigenvalues for the non-void region ($n_{str} > 1$) is shown. Total number of eigenvalue triplets are less than $10$ per cent of that of the full simulation box. Eigenvalues close to zero in non-void regions are notably fewer than in the entire simulation box.}
\label{fig:lambdasPDF}
\end{figure}

\begin{figure}
\begin{minipage}[t]{.99\linewidth}
  \centering\includegraphics[width=8.cm]{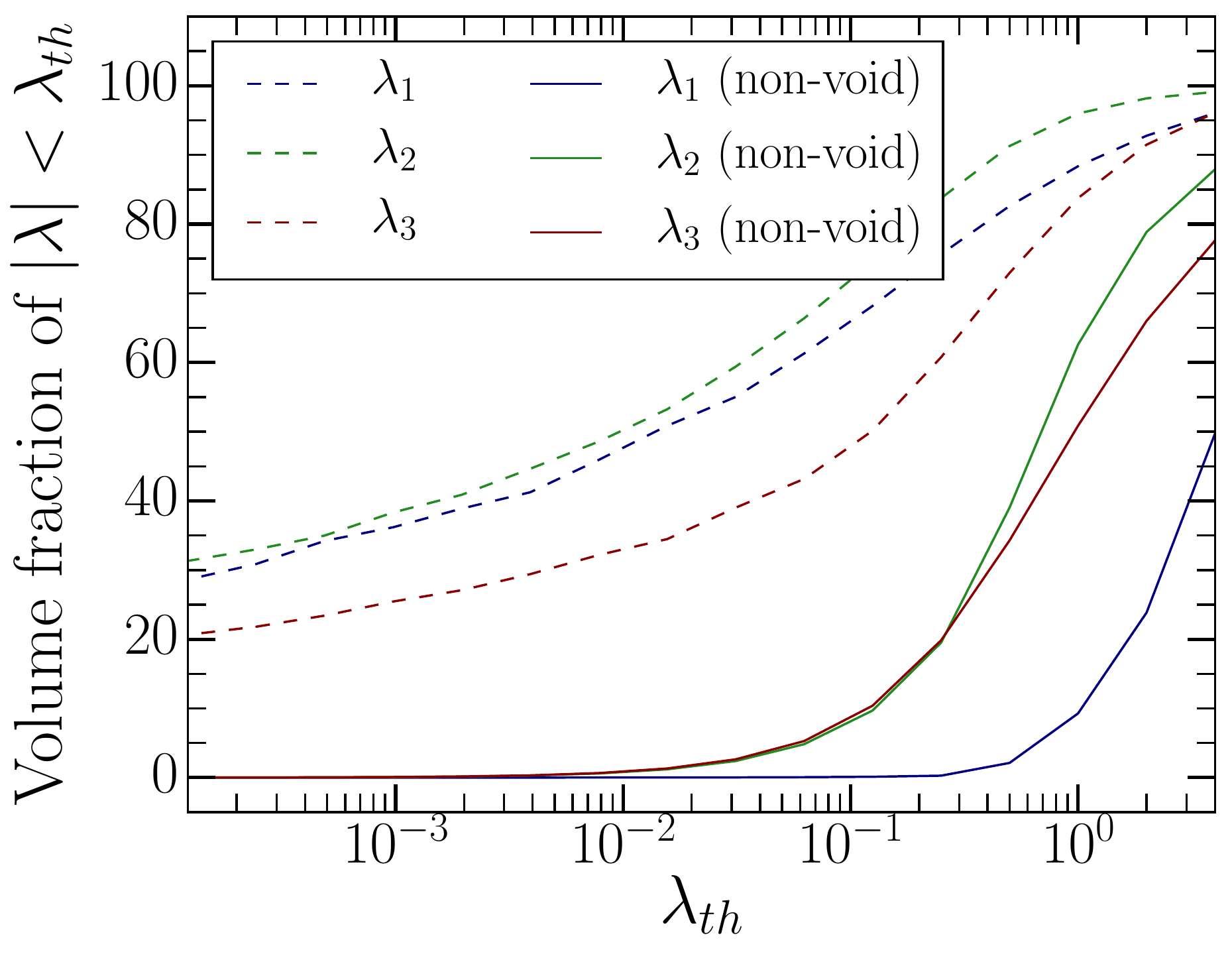} 

\end{minipage}\hfill
\caption{Comparison between small eigenvalues of the multistream Hessian $\mathbfss{H}(-n_{str})$. Percentage of eigenvalues with absolute values less than a cut-off, $\lambda_{th}$ are shown for full simulation box (dashed lines) and the multistream web structure (solid lines). The multistream web has fewer eigenvalues below $\lambda_{th} = 0.1$. The void seems to have most of the small eigenvalues. } 
\label{fig:lambdasSmall}
\end{figure}

The multistream field has a constant value of $1$ for around $90$ per cent of the simulation box. At least one gravitational collapse occurs in the remaining $10$ per cent of the volume. In these non-void regions, the $n_{str}$ value varies from 3 to very high values, often in the order of thousands. In the multistream field of refinement factor of 2 for simulation with $N_p = 128^3$ particles, maximum $n_{str}$ is 2831. Within the non-void structure, the multistream field may have several local maxima, minima and saddles. Variation of $n_{str}$ is especially high inside halo boundaries, where the particles in their non-linear stage of evolution have undergone a large number of flip-flops.  

Local second order variation in a scalar field $f$ like the multistream field can by found using the Hessian matrix $\mathbfss{H}(f)$. An element of the Hessian matrix is given in \autoref{eq:Hess}, where $i$ and $j$ can be any of $x$, $y$ or $z$ directions.  

\begin{equation}
\label{eq:Hess}
\mathbfss{H}_{ij}(f) = \frac{\partial^2 f}{\partial x_i \partial x_j} 
\end{equation}

In our analysis, we have chosen $f = -n_{str}(\bmath{x})$ for understanding local variations of the multistream field. The resulting Hessians at each point on the configuration space are always symmetric matrices, as illustrated in Appendix \ref{appendix:Eigen}. The eigenvalues of these Hessian matrices are always real, and depending on if their values are positive or negative, one may infer local geometrical features in the multistream field. 

Within the void, there is no variation in the multistream values. Hessians $\mathbfss{H}(-n_{str})$ are zero matrices in large volume fraction of the simulation box (around 90 per cent in both the simulations) due to the constant value of $n_{str} = 1$ in this percolating void. Eigenvalues of these Hessian matrices, sorted as $ \lambda_1 \geq \lambda_2 \geq \lambda_3 $ are close to 0 at a large number of regions as shown in the top panel of \autoref{fig:lambdasPDF}. In the simulation with $128^3$ particles, the median values of each eigenvalue are $0.09$, $-3\times 10^{-10}$ and $-0.11$ for  $\lambda_1$, $\lambda_2$ and $\lambda_3$ respectively. By selecting just the non-void region by $n_{str} > 1$, notably fewer number of eigenvalues have small absolute values. The median values of each of the eigenvalues in the non-void regions are $4.01$, $0.48$, and $-0.85$ respectively for $\lambda_1$, $\lambda_2$ and  $\lambda_3$. Bottom panel in \autoref{fig:lambdasPDF} shows a significant change in the probability distribution of Hessian eigenvalues around 0, the distribution pattern at the tails are mostly identical to the distribution pattern in the entire simulation box.

A large fraction of eigenvalues in non-void regions are still around 0, but their percentage is quite less compared to that of the entire box. For instance, nearly 66 per cent of $\lambda_1$'s, 72 per cent of $\lambda_2$'s and 48 per cent of $\lambda_3$'s are  within in the range of $0.0 \pm 0.1$ in the entire simulation box. However, with the exclusion of void regions, these volume fractions drops to 0.1, 7.7 and 8.4 per cent respectively (\autoref{fig:lambdasSmall}). Hence most of the eigenvalues at the void region have small absolute values.

Hessian eigenvalues in multistream fields differ from that in density, gravitational potential or velocity shear tensor. Constant scalar value of $n_{str}$ facilitates the Hessian $\mathbfss{H}(-n_{str})$ matrices to be presumptively close to zero. On the other hand, in density field manifests in a range of low values in the voids, resulting in non-zero Hessian matrices. Eigenvalues of velocity shear tensor do not peak at zero either \cite{Libeskind2013}. For the deformation tensor, morphological characterization of the cosmic web using Zel'dovich formalism shows that each eigenvalue must be negative in voids. 

\begin{figure*}
\begin{minipage}[t]{0.99\linewidth}
 \centering\includegraphics[height=14.cm]{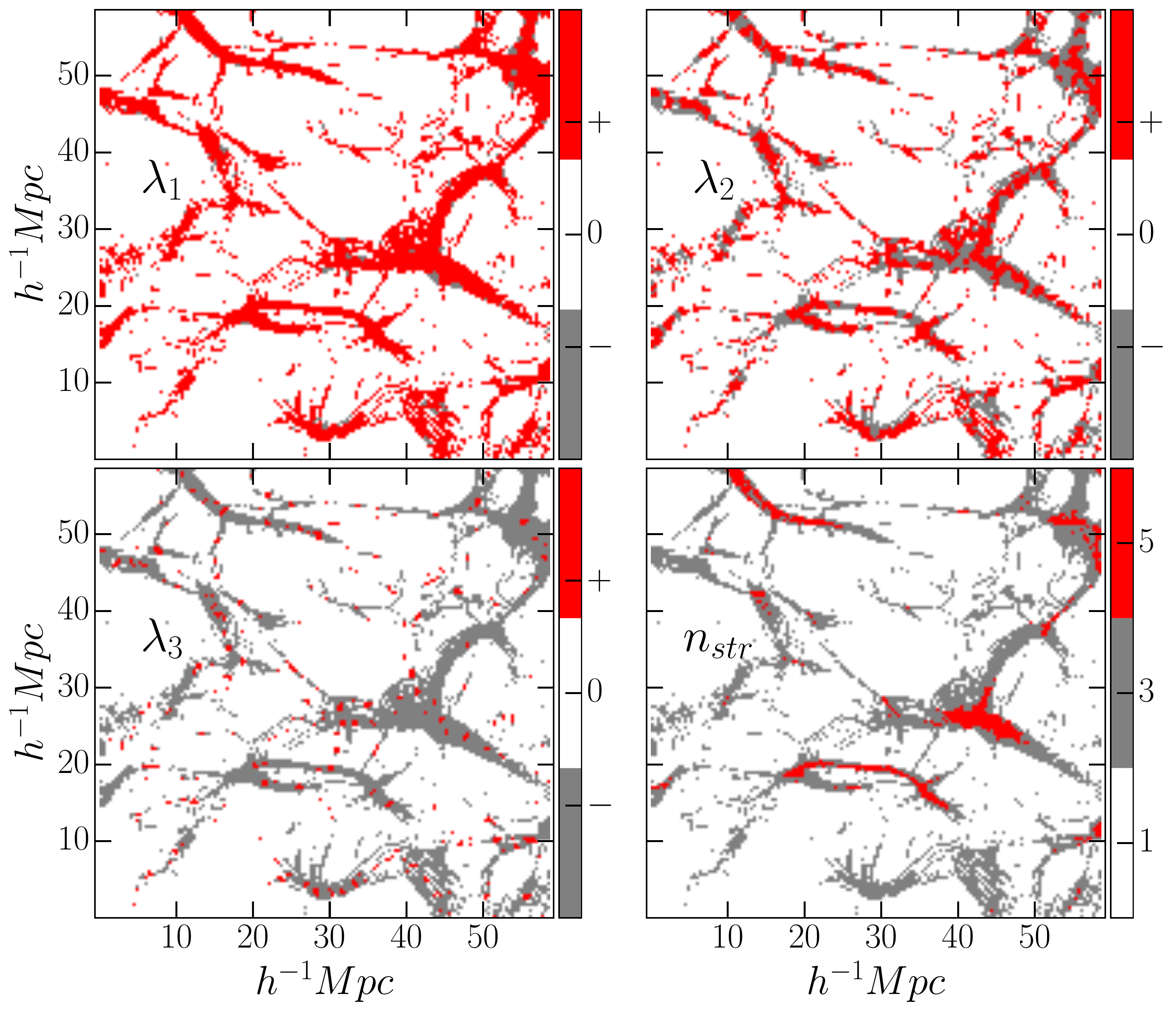} 
\end{minipage}\hfill
\caption{Eigenvalues of the Hessian matrix $\mathbfss{H}(-n_{str})$ in a slice of 50 $h^{-1}$ Mpc $\times$ 50 $h^{-1}$ Mpc slice of the simulation box of $128^3$ particles. Variation in the eigenvalues in the multistreaming web structure is shown. The largest eigenvalue $\lambda_1$ (top left panel) has positive values throughout the structure. The smallest eigenvalue $\lambda_3$ (bottom left) has negative values surrounding positive definite regions of the $n_{str}$ field. Corresponding multistream field is shown in the bottom right panel for single, three and more than five streams.}
\label{fig:evals123}
\end{figure*}

\begin{figure}
\begin{minipage}[t]{.99\linewidth}
  \centering\includegraphics[width=6.cm]{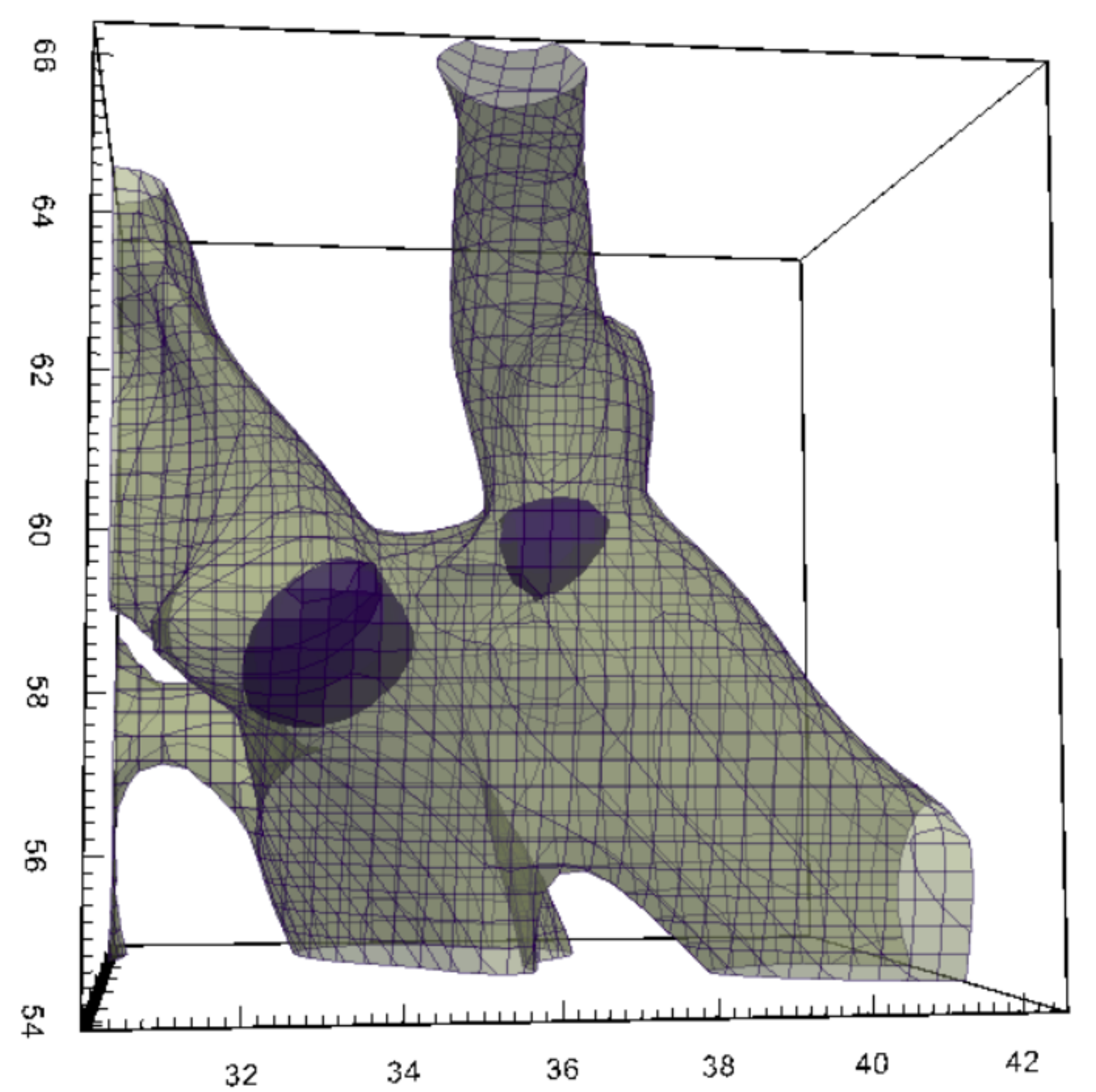} 
\end{minipage}\hfill
\caption{Surfaces identified in the multistream field. Blue regions are closed regions with $\lambda_3 > 0 $, which we identify as two haloes. Other surface has an open curvature along one direction, with $\lambda_1 > \lambda_2 > 0$ and $ \lambda_3 < 0$. }
\label{fig:SmallBox}
\end{figure}

The eigenvalues of $\mathbfss{H}(-n_{str})$ span a large range of values in our cosmological simulation. The largest eigenvalue of the triplets, $\lambda_1$ having large positive values throughout the multistream web structure ( see \autoref{fig:evals123} ). Absolute values $|\lambda_1|$, $|\lambda_2|$ and $|\lambda_3|$ peak around the neighbourhood of intersections of filaments. These junctions are usually high streaming regions due to shell crossing from multiple directions. \cite{Ramachandra2015} observed that these regions with intersecting filaments are in the vicinity of large FOF haloes.

If the Hessian matrices are positive definite in a region, i.e., if all the eigenvalues are strictly positive, then the interior of this convex region has at-most one minimum. For our choice of $-n_{str}(\bmath{x})$ as the domain of Hessian, this means that the convex neighbourhoods around local maxima of the multistream field are isolated by the positive definite Hessian matrices. Closed surface contours at high streaming or the most non-linear regions are selected. These regions my indeed be the regions of dark matter haloes.

The smallest eigenvalue, $\lambda_3$ has lowest volume fraction of all the eigenvalues in the positive tail of the distributions in \autoref{fig:lambdasPDF}. Since the condition $\lambda_3 > 0$ ensures the Hessian matrix to be positive definitive, we may use it as a primary criterion in isolating compact regions of dark matter haloes. These regions also roughly correspond to isolated globs as seen in \autoref{fig:SmallBox}. Local geometry analysis is pertinent for halo detection due to compact geometry of the haloes. In principle, other components of the cosmic web could also be detected. Tubular structures in filaments could be detected, as shown in \autoref{fig:SmallBox}, using conditions on the eigenvalues as $\lambda_1 > \lambda_2 > 0$ and $ \lambda_3 < 0$ . Fabric-thin walls could be detected by $\lambda_1 > 0$ and $ \lambda_3 < \lambda_2 < 0$. 

\subsection{Softening of the multistream field}
\label{sub:Softening}


Hessian eigenvalues are generally defined on continuous functions. Although our domain of the Hessian is an inhererntly integer-valued field, it describes the multistream structure at the level of diagnostic grid. Hence it may be considered to be numerically equivalent to a continuous function where the numerical approxiamtion of differentation is a valid operation. This can be verified mathematically by finding that Hessian $\mathbfss{H}(-n_{str})$ is symmetric (Appendix \ref{appendix:Eigen} shows the numerical approximation of the Hessian matrix term for generic unfiltered multistream field.)

\begin{figure}
\begin{minipage}[t]{.99\linewidth}
  \centering\includegraphics[width=8.cm]{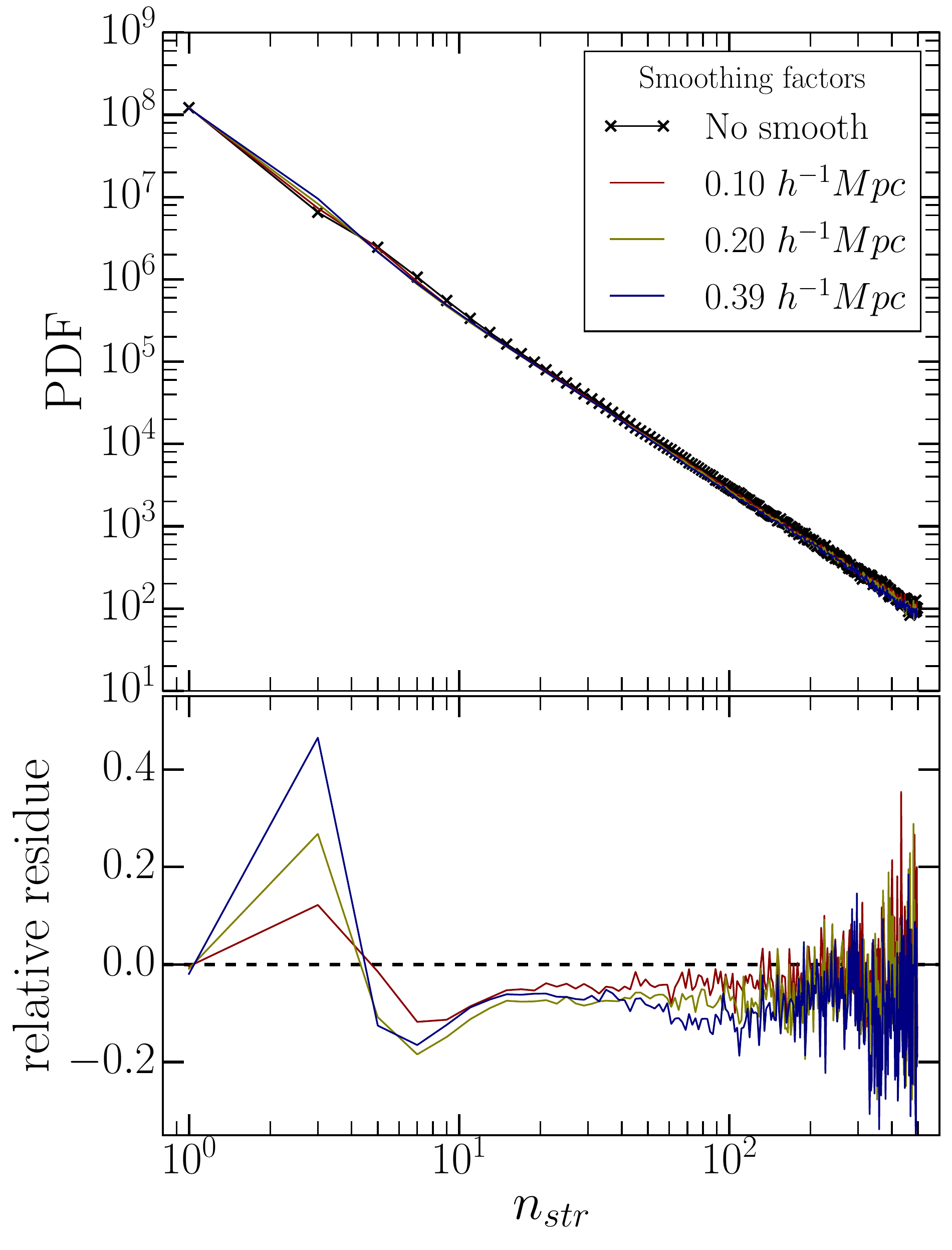} 

\end{minipage}\hfill
\caption{Probability distribution function of the multistream $n_{str}$ values in the simulation box with $N_p = 256^3$. The multistream field is calculated at refinement factor $l_l/l_d= 2$. Unsmoothed multistream field is compared with different Gaussian filtering scales. Softening scales of equal to $0.5$, $1$, and $2$ times the side length of diagnostic grid $l_d$ correspond to $0.10 h^{-1} \text{Mpc}$, $0.20 h^{-1} \text{Mpc}$, and $0.39 h^{-1} \text{Mpc}$ respectively.}
\label{fig:nstrSmooth}
\end{figure}

\begin{figure}
\begin{minipage}[t]{.99\linewidth}
  \centering\includegraphics[width=8.cm]{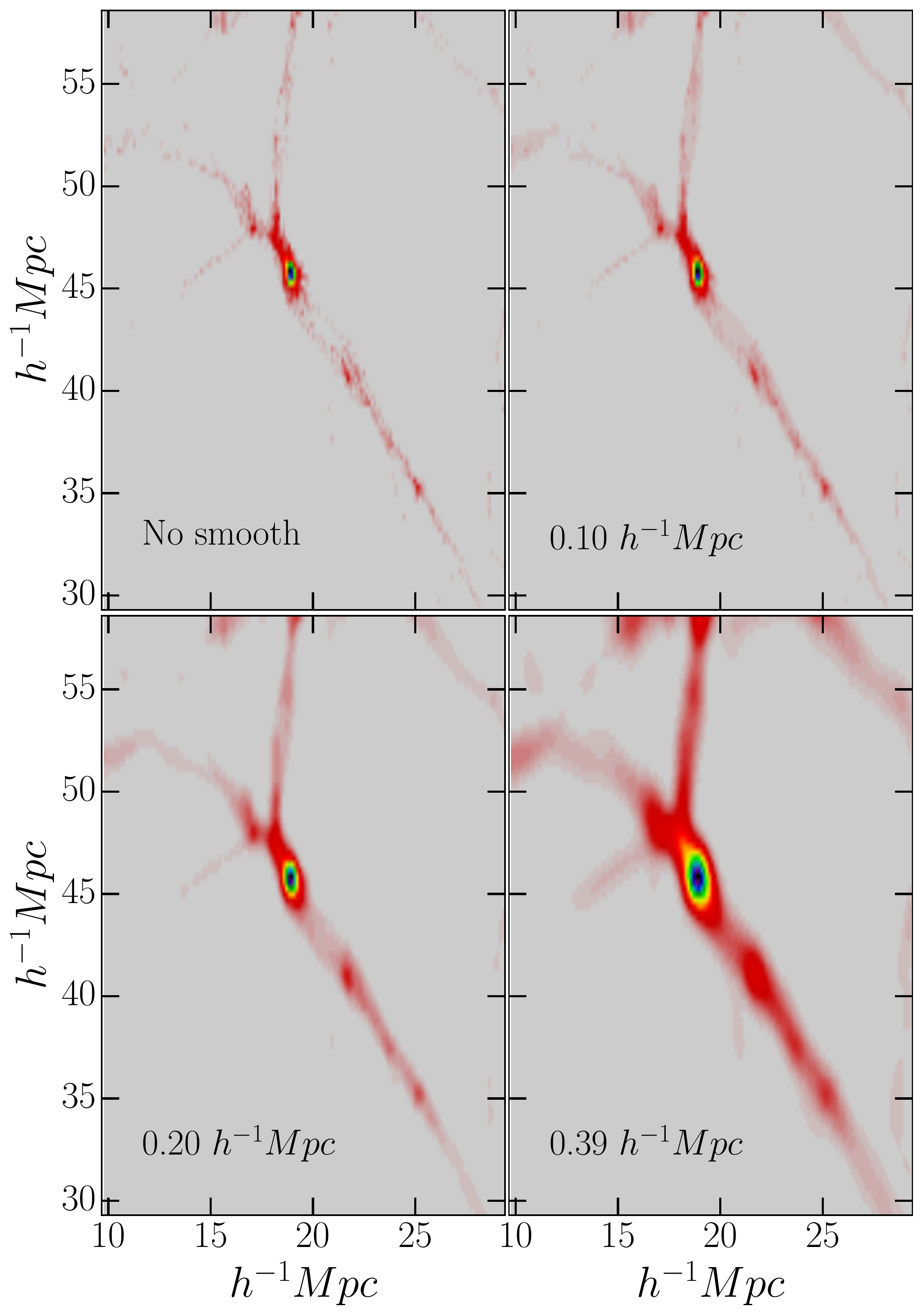} 

\end{minipage}\hfill
\caption{Multi-stream field at various softening scales in the simulation box with $N_p = 256^3$. The multistream field is calculated at refinement factor $l_l/l_d= 2$. Unsmoothed multistream field is compared with different Gaussian filtering scales equal to $0.10 h^{-1} \text{Mpc}$, $0.20 h^{-1} \text{Mpc}$, and $0.39 h^{-1} \text{Mpc}$ respectively.}
\label{fig:nstrSmoothSmall}
\end{figure}

Smoothing the multistream field (at the refinement level of $l_l/l_d=$ $1$ or $2$) effectively reduces noise. There is also a systematic variation in the distribution of smoothed $n_{str}$ values as shown in \autoref{fig:nstrSmooth}. Volume fraction of the single-streaming voids only varies from $90.8$ per cent without smoothing to $89.1$ per cent for the Gaussian softening length of 0.39 $h^{-1} Mpc$ (twice the length of diagnostic grid $l_d$). On the other hand, $n_{str} = 3$ regions gain volume fraction from $4.9$ per cent in un-smoothed field to $7.1$ per cent for 0.39 $h^{-1} Mpc$. This is seen in the multistream structures of smootheing scales of 0.39 $h^{-1} Mpc$ in \autoref{fig:nstrSmoothSmall}. Multi-stream regions with $ 3 < n_{str} \leq 100 $ occupy correspondingly lower volumes for higher smoothing, and the variation is noisy beyond $n_{str} > 100$. \autoref{fig:nstrSmoothSmall} shows the multistream field on a small slice of the simulation at different softening scales, and walls and filaments are resolved better with increasing softening.                                                                

Smoother multistream fields result in less noisy PDFs of the Hessian eigenvalues. For instance, the volume fraction of regions with positive curvature (i.e. $\lambda_3 > 0$) is $2.4\%$, $2.3\%$ and $2.5\%$ for scales $0.20 h^{-1} \text{ Mpc}$, $0.39 h^{-1} \text{ Mpc}$, $0.78 h^{-1} \text{ Mpc}$ respectively. Further analysis of smoothed positive definite regions is relevant in determining halo boundaries, and will be extensively discussed in the next paper. 

\subsection{Resolution dependence}
\label{sub:resolution}

Multi-stream calculation can be done at arbitrarily high resolutions by populating the tetrahedral simplices. For our resolution study, we have chosen a smaller slice of $ 50 h^{-1} \text{Mpc} \times 50 h^{-1} \text{Mpc} \times 50 h^{-1} \text{Mpc}$ (grid of size $64^3$ from the N-body simulation) from the simulation with $N_p = 128^3$ particles. The multistream field is calculated at 4 different refinement factors, i.e., at diagnosis grids of size $64^3$ ($l_l/l_d = 1$), $128^3$ ($l_l/l_d = 2$), $256^3$ ($l_l/l_d = 4$)  and $512^3$ ($l_l/l_d = 8$) respectively.

Volume fractions of each multistream does not change systematically for different levels of refinement, except at very high $n_{str}$ values (see \citealt{Ramachandra2015} for dependence of $n_{str}$ variation on refinement of the diagnostic grid). At high multistream values, higher resolutions reveal a considerably less noisy multistream fields. 

There are no variations in the volume fractions of the cosmic web components classified using the global $n_{str}$ thresholds as shown in \autoref{tab:VolfrRefNst}. Voids ($n_{str} = 1$) occupy about 90 per cent of the volume at each refinement factor. Rest of the heuristic thresholds that identify the structure components (as prescribed by \citealt{Ramachandra2015}) are constant multistream contours: $3 \leq n_{str} < 17$ for walls, $17 \leq n_{str} < 90 $ for filaments and $n_{str} \geq 90 $ for haloes. Since the volume fraction of each $n_{str}$ values are about the same at each refinement factor, the volume fraction of the cosmic web components corresponding to global multistream thresholds do not vary considerably.

\begin{table}
\caption{Volume fraction (in per cent) of $n_{str}$ thresholds for cosmic web structures as defined by \protect\cite{Ramachandra2015}. Multi-stream field is calculated at 1, 2, 4, and 8 times the native simulation resolution of $64^3$ grids. Small slice of $ 50 h^{-1} \text{Mpc} \times 50 h^{-1} \text{Mpc} \times 50 h^{-1} \text{Mpc} $ is chosen for the analysis.}
\begin{tabular}{|l|l|l|l|l}
\hline
Global thresholds  & $64^3$   & $128^3$   & $256^3$  & $512^3$ \\  \hline
$n_{str} = 1$ (Void)     &    90.87   & 90.92  & 90.94 & 90.94 \\ \hline
$3 \leq n_{str} < 17$ (Wall)  & 8.71 & 8.66 & 8.63 & 8.64   \\ \hline
$17 \leq n_{str} < 90 $ (Filaments) &  0.39 & 0.39 & 0.39 & 0.39    \\ \hline
$n_{str} \geq 90 $ (Haloes) & 0.034 & 0.035 & 0.036 & 0.036 \\ \hline

\end{tabular}
\label{tab:VolfrRefNst}
\end{table}

However, local geometry analysis of the multistream flow field varies considerably on the resolution of the analysis grid. For our Hessian $\mathbfss{H}(-n_{str})$, the regions with $\lambda_1 \geq \lambda_2 \geq \lambda_3 > 0$ in non-void regions occupy 1.8 per cent of the entire box in native resolution of diagnostic grid, as shown in \autoref{tab:VolfrRefLambda}. This fraction reduces to 1.3 per cent at diagnostic grid of $512^3$ resolution. Variations with refinement factors are seen in other eigenvalue conditions in the non-void too: volume fraction of $\lambda_1 > 0 > \lambda_2 \geq \lambda_3 $ regions increases from 1.7 per cent at refinement factor of 1 to 3 per cent at refinement factor of 8. Volume fraction of $\lambda_1 \geq \lambda_2 > 0 > \lambda_3 $ regions decreases from 5.6 to 4.6 per cent with the increase of refinement from 1 to 8. 

In principle, the conditions for geometric criteria are: $\lambda_1 > 0 > \lambda_2 \geq \lambda_3$ for locally flat regions, $\lambda_1 \geq \lambda_2 > 0 > \lambda_3 $ for locally tubular structures and $\lambda_1 \geq \lambda_2 \geq \lambda_3 > 0$ for clumped blobs. However, the tabulated the volume fractions in \autoref{tab:VolfrRefLambda} does not correspond to cosmic web components themselves. Identification of the components may require post processing steps. 

High resolution studies of multistream fields would play an important role in detection of walls and filaments. These two components have smaller length scales along at least one direction with respect to others. As seen in Section \ref{sec:voidPerc}, walls are more resolved in high resolution of multistream fields, enclosing pockets of voids (see \autoref{fig:voidFace}).  

However, a Hessian analysis to identify filaments and walls may be considerably different from that of halo finding due to the following reasons: First, a local geometrical analysis is uniquely convenient for detecting dark matter haloes since they are local structures. Filaments and walls, alternatively, are structures that span over large distances. Secondly, we try to find regions around local maxima of multistream field for haloes. Whereas, filaments and walls have much weaker relationship with local multistream maxima. Filaments and walls usually deviate from flat planar or straight tubular geometries: they often have complicated structures several connections and branches. For these reasons, Hessian eigenvalues alone would not be sufficient in detecting walls or filaments. 

\begin{table}
  \caption{Volume fraction of criteria based on $n_{str}$ and $\lambda$s of $\mathbfss{H}(-n_{str})$ calculated at various resolutions. We chose a smaller slice of $ 50 h^{-1} \text{Mpc} \times 50 h^{-1} \text{Mpc} \times 50 h^{-1} \text{Mpc} $ i.e., half the volume of the original GADGET simulation box. The refinement factors are the multiplication factors of 1, 2, 4 and 8 times of the native resolution ($64^3$) of the simulation grid along each axis. Eigenvalues of the Hessian of the field are local geometric parameters. The void is globally defined as $n_{str} = 1$ and the multistream web structure as $n_{str} > 1$.}

\begin{tabular}{|l|l|l|l|l}
\hline
Global/local conditions & $64^3$   & $128^3$   & $256^3$  & $512^3$  \\  \hline
$n_{str} = 1$ (Void)     &    90.87   & 90.92  & 90.94 & 90.94 \\ \hline
$n_{str} > 1$; $\lambda_1 > 0 > \lambda_2 \geq \lambda_3 $  &  1.72 & 2.22 & 2.67 & 2.96    \\ \hline
$n_{str} > 1$; $\lambda_1 \geq \lambda_2 > 0 > \lambda_3 $ & 5.60 & 5.28 & 4.91 & 4.57   \\ \hline
$n_{str} > 1$; $\lambda_1 \geq \lambda_2 \geq \lambda_3 > 0$ & 1.81 & 1.56 & 1.37 & 1.26 \\ \hline

\end{tabular}
\label{tab:VolfrRefLambda}
\end{table}

\section{Discussion}
\label{sec:discussion}

Formation of multiple velocity streams in the context of structure formation has been known in the past, starting from Zel'dovich approximation. Quantification of the multistreams in N-body simulations, however, was recently achieved by \cite{Shandarin2012} and \cite{Abel2012b} using the Lagrangian sub-manifold. In our study, the multistream fields are calculated using the tessellation algorithm by \cite{Shandarin2012}. We have analysed, for the first time, the local geometry and percolation properties of the cosmic web using this multistream field.   

Distinguishing the configuration space into void and non-void is one of the uses of the multistream field. Lagrangian sub-manifold has no folds in the beginning, thus $n_{str} = 1$ uniformly throughout the simulation. Gravitational instability folds the sub-manifold in complicated ways, however, most of the volume has particles without any collapse. \cite{Shandarin2012} and \cite{Ramachandra2015} observed that the single-streaming voids occupy around 85-90 per cent of the simulations at $z=0$. In this study, we found that the void regions are also connected in a way that the largest percolating void occupies more than 99 per cent of the all the single-streaming regions. Recent study by \cite{Wojtak2016a} uses a watershed transform method in the density field prescribed by Lagrangian tessellations (\citealt{Shandarin2012} and \citealt{Abel2012b}) to analyse the evolution of isolated voids. Another recent study by \cite{Falck2015} on ORIGAMI-voids also reveal a similar percolation at the limit of simulation resolution. They observed persistence of this phenomenon for different resolutions of the N-body simulation. Multi-stream analysis, on the other hand, is not limited to mass resolution of the simulation. Our multistream analysis refined upto 8 times the simulations resolution revealed that the percolation phenomenon still persists. However, at high refinements of the multistream field, we observed small voids that are enclosed by highly resolved non-void membranes. 

 
\begin{figure}
\begin{minipage}[t]{.99\linewidth}
  \centering\includegraphics[width=10.cm]{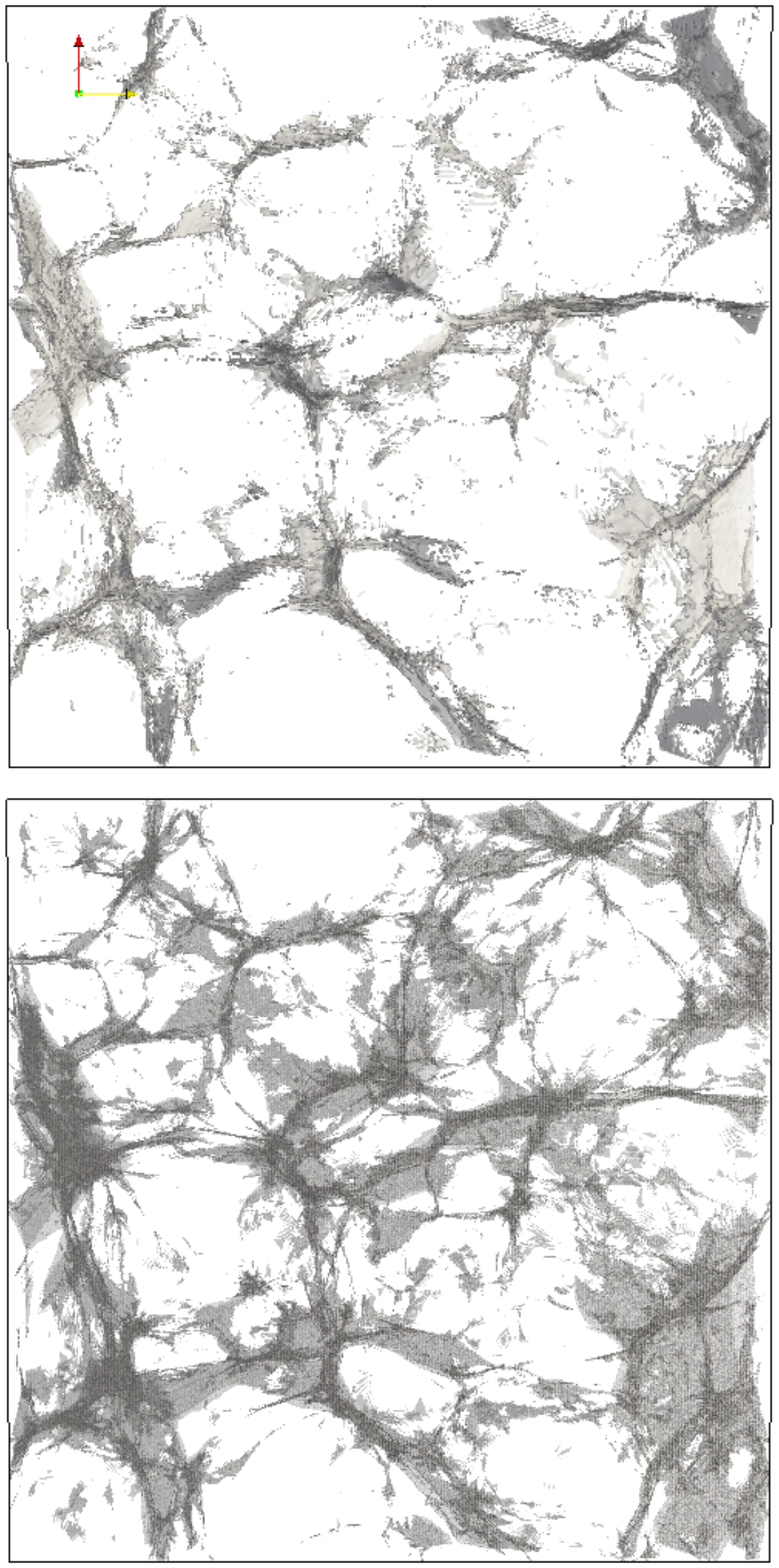}
  \centering\includegraphics[width=10.cm]{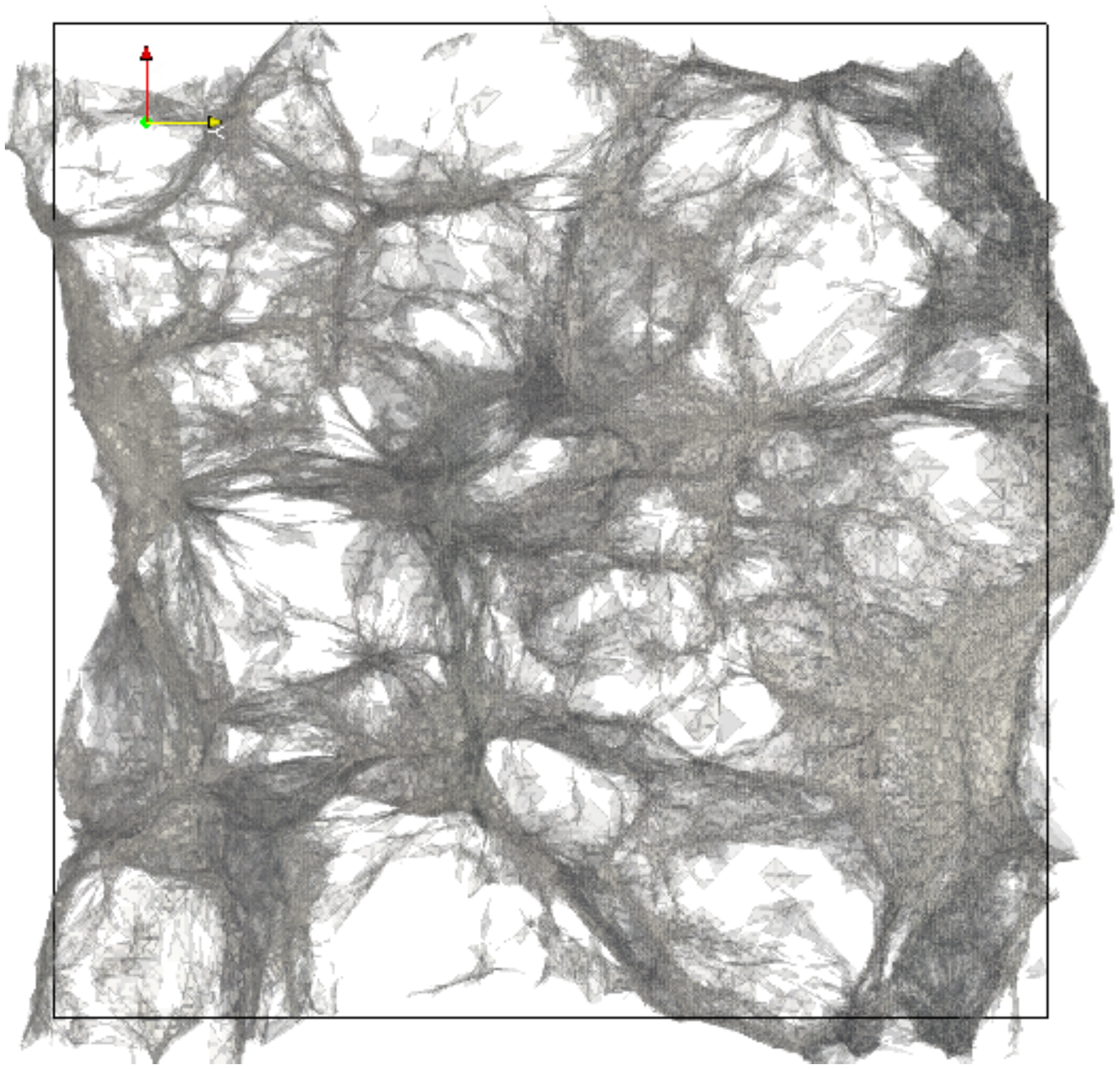}  
\end{minipage}\hfill
\caption{Two top panels show  three contours ($n_{str}=3, 11, 17$) in a slice $100h^{-1} \text{Mpc} \times 100h^{-1} \text{Mpc} \times 10h^{-1} \text{Mpc}$ in the simulation with $128^3$ particles, computed at two refinement factors: 2(upper) and 8 (lower). The bottom panel shows the caustic surfaces in the same slice. }
\label{fig:NstrCaust}
\end{figure}

Walls are the first collapsed structures in the dark matter Universe. At highly refined multistream field, thin membranes of the structures are often resolved, revealing small voids enclosed by them (compare two top panels in \autoref{fig:NstrCaust}). These preliminary structures are separated from the voids by caustic surfaces. These caustics have volume measure zero, which makes detection of their surface harder in the multistream field, even at very high resolutions. On the other hand, caustic surfaces themselves can by detected using the Lagrangian sub-manifold by identifying the common faces of neighbouring tetrahedra 
with opposite volume signs \citep{Shandarin2012}. They are shown in the bottom panel in  \autoref{fig:NstrCaust}. One can see that increasing the refinement factor from 2 to 8 adds mostly walls but the complete wall structure shown in the bottom panel is still considerably greater. Please note that the plots in  two top panels adjusted  exactly to the simulation box in Eulerian space, and the bottom plot shows the Lagrangian box mapped to Eulerian space without adjusting to the simulation box.

There are extensive number of topological indicators in the context of density fields or spatial co-ordinates - such as alpha shapes, Betti numbers, genus statistics. Although a comparitive study of these topolgical measures in multistream fields may be interesing, it is not the intent of this paper. In this study, we only investigate percolation transitions in excursion sets of multistreams as a preliminary analysis of topological connectivities. Excursion sets in density fields are shown to have quick percolation transitions \citep{Shandarin2010b} and a similar trend in multistream field is investigated here.

Excursion sets of multistream and density field (calculated using CIC and DTFE in this study) reveal some of the topological differences. At any volume fraction of excursion set $f_{ES}$, the filling factor of the largest structure $f_{1}/f_{ES}$ is lower for mass density (both CIC and DTFE). This concludes that the mass density field is more fragmented than the multistream field. A large number of disconnected segments are seen at high $n_{str}$ or $\rho/ \rho_b$ thresholds, and the number of connections increase with decreasing $n_{str}$ threshold.

Global connectivities in the cosmic web is slightly different for multistream field and the density field. The largest structure in the excursion set starts percolating at certain values of excursion volume fraction ($f_{ES}$). As shown in Section \ref{sec:percolation}, these percolation transitions occur at $\rho_{DTFE}/ \rho_b = 5.16 $, $\rho_{CIC}/ \rho_b = 5.49 $ for density fields and $n_{str} = 17$ for the multistream field. The corresponding percolation volume fraction $f_{ES}^{(p)}$ is smaller for multistream fields ( $f_{ES}^{(p)} = 0.75$  per cent for multistream field and $f_{ES}^{(p)} = 1.7$ per cent for the CIC-density field $f_{ES}^{(p)} = 2.9$ per cent for the DTFE-density field). This indicates that the percolating multistream filament is over 2 times thinner than that of $\rho_{DTFE}$ and over 3 times thinner than $\rho_{CIC}$ field. 

Since the $n_{str}$ field in this study is calculated on regular grids, the boudaries of the structures are not exactly traced. Outlining foldings in the Lagrangian sub-manifolds exactly as shown in \autoref{fig:NstrCaust} or in the {\it flip-flop} calculations shown in \cite{Shandarin2016} give point datasets which are considerably more difficult to analyze. However, recent advancements in computational topology - such as the adaptation of the watershed transforms (using SpineWeb -\citealt{Aragon-Calvo2008} and Morse theory (using DisPerSe - \citealt{Sousbie2011e} and Felix - \citealt{Shivshankar2015a}) to inherently discrete datasets may be useful in the topological analyses of flip-flop fields and caustics.

The multistream field is a scalar function of Eulerian coordinates. We have analysed functional variation of the $-n_{str}(\bmath{x})$ field using Hessian eigenvalues. The Hessian analysis is generally done for inherently continuous fields, For example, Hessian analysis has been previously studied for smoothed density fields (see \citealt{Sousbie2008a}, \citealt{Aragon-Calvo2007}, \citealt{Aragon-Calvo2010}, \citealt{Cautun2014a} etc.), gravitational potential and velocity shear tensor (\citealt{Hoffman2012a}, \citealt{Libeskind2013}, \citealt{Hahn2007}, \citealt{Forero-Romero2009a}, \citealt{Hoffman2012a} and \citealt{Cautun2014a}). Although the multistream field has discrete values by definition, it may be considered smooth for numerical analysis at the scale of grid length of the field. The resulting Hessian eigenvalues characterize the geometry in a four-dimensional hyper-space of ($-n_{str}, x, y, z$). The boundary of a region with $\lambda_1 \geq \lambda_2 \geq \lambda_3 > 0$ is a closed convex contour in this hyper-space, and thus it's projection onto the three-dimensional Lagrangian space is also closed and convex.

Dark matter haloes, being localised structures, are uniquely convenient for our local Hessian analysis. Conditions of $\lambda_1 > 0 > \lambda_2 \geq \lambda_3 $ and $\lambda_1 \geq \lambda_2 > 0 > \lambda_3 $ also give information about curvature. Hessian eigenvalue analysis at high resolution of multistream fields may be very interesting in understanding the tubular edges of filaments and surfaces of walls at smaller scales. 

\section{Summary}
\label{sec:summary}

We studied certain geometrical and topological aspects of the multistream field in the context of large scale structure of the Universe. Several features were found to be considerably different from traditional density fields. The major findings from our analysis are briefly summarized as follows: 

\begin{enumerate}

\item We use the multistream field as a proxy for distinguishing of the DM web from DM voids: the web is defined as the regions
with number of streams greater than one and thus voids  as  a single stream regions. The boundary between them  
representing a sharp transition from one-- to three-- stream flow regions
would be a caustic surface in the density field if the mass and spatial resolutions were sufficiently high. They were clearly seen in 
2D simulations by \cite{Melott1989} as well as in 3D simulations by \cite{Angulo2016}, \cite{Hahn2016a}, \cite{Hahn2013} and in velocity fields \cite{Hahn2015a}.

\item Regions without any folds in the Lagrangian sub-manifold are mostly connected. These single streaming void regions at $z=0$ occupy around 90 per cent of both simulations used in this study, most of which belong to a single percolating structure. However at high resolution multistream analysis, we identify a number of isolated pockets that are entirely enclosed by boundary of walls. But these voids are tiny and collectively occupy less than 0.1 per cent of the volume of the simulation box.

\item The Hessian components of the multistream field are universally zero in the interior of the void, due to constant value of $n_{str}$. Density field need not have zero Hessians since mass density is not unequivocally constant at $z=0$.

\item We studied the global topology of the non-void ($n_{str} > 1$) structure using percolation analysis. A rapid percolation transition occurred in our multistream field at $n_{str} = 17$. The percolating filament in multistream field is thinner than the percolating filament in mass density field.

\end{enumerate}

The Lagrangian sub-manifold contains dynamical information of structure formation. We analysed the multistream field that contains the information of foldings in the sub-manifold.  Connectivities in the void and non-void components of the multistream web reveal several details about structure of the Universe that are not probed by traditional density fields. In addition, we demonstrated the use of geometrical features of the multistream field in identifying potential dark matter halo candidates in cosmological N-body simulations.

\section*{Acknowledgements}

This work has been funded in part by the University of Kansas FY 2017 Competition General Research Fund, GRF Award 2301155. This research used resources of the Advanced Computing Facility, which is part of the Center for Research Computing at the University of Kansas. We thank Mark Neyrinck and Mikhail Medvedev for discussions and suggestions. We also thank the anonymous referee for insightful comments on improving this manuscript. 

\bibliographystyle{mnras}
\bibliography{library}

\appendix

\section{The multistream flow field in one-dimension}
\label{appendix:nstream}

\begin{figure}
\begin{minipage}[t]{.99\linewidth}
  \centering\includegraphics[width=8.cm]{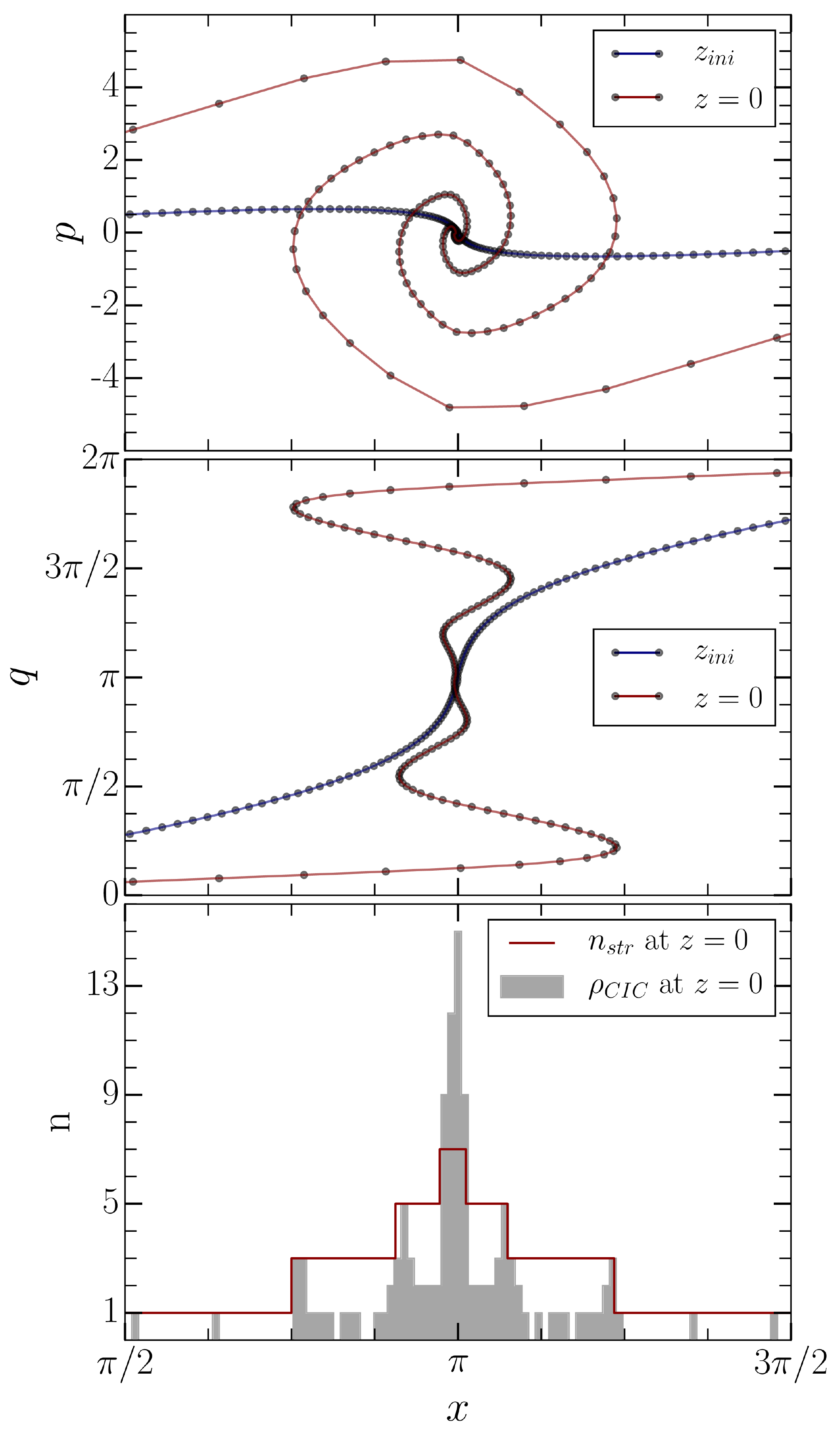} 
\end{minipage}\hfill
\caption{ Multi-streaming in one-dimension gravitational collapse. Top panel: $(\bmath{p}, \bmath{x})$ phase-space representation redshift $z_{ini}$ and $z = 0$. Dots represent the dark matter particles. Initially the mass particles are in the linear stage of evolution. At $z = 0$, multiple values of $\bmath{p}(\bmath{x})$ is seen in the collapsed regions. Middle panel: Equivalent Lagrangian sub-manifold $\bmath{q}(\bmath{x})$. At $z_{ini}$, the dashed line represents $\bmath{q} = \bmath{x}$. Number of streams are parametrized from this sub-manifold. Bottom panel: The multistream field $n_{str}$ and the number-density using CIC algorithm, $n_{CIC}$ at $z = 0$. }
\label{fig:phase1d}
\end{figure}

The top panel in \autoref{fig:phase1d} shows the velocity multistreaming phenomenon in a one-dimensional collapse. The phase-space $(\bmath{p}, \bmath{x})$ (where $p$ is the momentum and $x$ is the co-moving Eulerian coordinate) is single-valued in the linear stage of evolution (at redshift $z_{ini}$). Non-linear stage of gravitational evolution of the collision-less dark matter particles then results in multi-valued $\bmath{p} (\bmath{x},z)$ at $z = 0$. The mass particles are sparsely  distributed outside the region of gravitational collapse, and are denser in the inner streams.

A dynamically equivalent transformation $(\bmath{p}, \bmath{x}) \mapsto (\bmath{q}, \bmath{x}) $ (where $\bmath{q}$ is the Lagrangian coordinate) shows the Lagrangian sub-manifold in the middle panel of \autoref{fig:phase1d}. This two-dimensional phase-space has foldings that correspond to multiple velocity streams, although the sub-manifold itself remains continuous. A projection of the Lagrangian sub-manifold at each point in the configuration space quantifies the number-of-streams. Folding in the sub-manifold are checked for points in configuration space using tessellations. The tessellating simplices in one-dimensional model are just the line-segments whose nodes are the dark matter particles in the Lagrangian space. Dynamical property is accounted for in this phase-space tessellation since labels of the nodes remain intact throughout the evolution; the line segments may shorten, extend or change orientation. Each folding in the Lagrangian sub-manifold increases the number of streams by a factor of two. In three-dimensional simulations, the sub-manifold twists in complicated ways in a six-dimensional phase space. The number-of-streams in N-body simulations (\citealt{Shandarin2012} and \citealt{Abel2012b}) is calculated using Lagrangian/phase-space tessellations. This triangulation is conceptually different from the Voronoi (See \citealt{Schaap2000} and references therein) or Delaunay \citep{Icke1991} tessellation schemes. 

The bottom panel \autoref{fig:phase1d} shows the the multistream field $n_{str}(\bmath{x})$ at $z = 0$. The field only takes the values of 1, 3, 5 and 7 in this scenario. Caustics occur at the folds in Lagrangian sub-manifold, and have a measure zero (study of caustics in one- and  two-dimensional evolution is done in \cite{Hidding2014}, three-dimensional caustic surface in a cosmological simulation is shown in \autoref{fig:NstrCaust}). Several properties of the multistream field are significantly different from mass density. The bottom panel also shows an illustration of CIC algorithm (cf. \citealt{Hockney1988}) in calculating density, which is numerically equivalent to counting the number of particles on each cell of a regular grid. One major difference is in the regions before gravitational collapse: $n_{str}$ is universally equal to unity, whereas number density fluctuates. It should also be noted that density by definition is a continuous field; numerical approximations like CIC discretise the field. Alternatively, multistream field is intrinsically a discrete-data field.

\section{Variations in the multistream field}
\label{appendix:Eigen}

A second-order local variations of a scalar field $f$ is described by a Hessian. In a three-dimensional domain, the Hessian is given by \autoref{eq:Hess}. The geometry of the scalar field is classified by the Eigenvalues of the Hessian. The convex regions have at-most one maxima within the (3+1)-dimensional functional space. Projection of this closed region onto three-dimensional coordinate space also gives a closed surface in coordinate space. 

We treat $n_{str}$ approximately continuous, for which the Hessian is always symmetric. In this study we use the scalar field $n_{str}(\bmath{x})$ inherently has discrete values like 1, 3, 5, and so on. The equation for numerical differentiation in the off-diagonal terms using Forward-difference method (using step-sizes of $\Delta x_i$ and $\Delta x_j$ along $i$ and $j$ respectively) is given in \autoref{eq:part1}. Notice that $\frac{\partial^2 f}{\partial x_i \partial x_j} = \frac{\partial^2 f}{\partial x_j \partial x_i}$, since RHS in \autoref{eq:part1} remains same. Hence the Hessian matrix in \autoref{eq:Hess} for the discrete scalar field $n_{str}$ is always numerically symmetric. Backward or central difference give similar results too. Smoothing of the multistream field further reduces any numerical noise in the Hessian eigenvalues.

\begin{equation}
\label{eq:part1}
\frac{\partial^2 f}{\partial x_i \partial x_j} = \frac{1}{\Delta x_i \Delta x_j}  \left[f_{i+1,j+1,k}-f_{i,j+1,k}-f_{i+1,j,k}+f_{i,j,k} \right]
\end{equation}

An integer-valued function, like the multistream field, is either constant or changes by a constant value in its real domain. In addition, the transitions in the multistream field are of multiples of 2, unless caustic surfaces are detected at the exact grid location. Consider $f_{i,j,k} = n$ at any grid point. Due to the property of multistream field, the values in the neighbourhood differ by a multiple of 2. That is,  $f_{i+1,j,k} = n+2p$, $f_{i,j+1,k} = n+2q$, $f_{i+1,j+1,k} = n+2r$, for some integers $p$, $q$ and $r$. Thus the second order variation of the multistream field reduces to \autoref{eq:part2}. 

\begin{equation}
\label{eq:part2}
\frac{\partial^2 f}{\partial x_i \partial x_j} = \frac{1}{\Delta x_i \Delta x_j}  \left[ 2r - 2p + 2q \right]
\end{equation}

Thus the numerical differentiation is independent of $n_{str}$ itself. It's important to note that this behaviour of the multistream field is independent of grid size. Also, the second order variation is a ratio of an even-number and the face area of the grid cube. The \autoref{eq:part2} becomes zero in a trivial case of $r = p = q = 0$, which corresponds to regions where $n_{str}$ is constant, including voids. In the non-trivial case, $r=(p+q)$, for non-zero $r$, $p$ and $q$. In the multistream grid, $2(p+q)$ could be considered as sum of variations in $n_{str}$ in the immediate neighbouring grid points. And $2r$ is the variation between next closest grid point, which is along the face-diagonal. 

On the other hand, mass density fields have sharp peaks at the multistream transitions. These peaks in the at the location of caustic are far less predictable, since the density fields become extremely noisy. For instance,\cite{Vogelsberger2011b} show noisy peaks of varying magnitude at the at high resolutions of mean density near halo locations. At lower resolutions, these sharp peaks are smoothed out, hence giving the impression of a smooth field. \cite{Hahn2015a} show similar `ill-behaved' derivatives in velocity fields at the caustic locations, where the derivatives are infinite.

\label{lastpage}

\end{document}